\documentclass[aps,preprint,prd,superscriptaddress,nofootinbib]{revtex4-2}

\usepackage{graphicx}  % needed for figures
\usepackage{dcolumn}   % needed for some tables
\usepackage{bm}        % for math
\usepackage{amssymb}   % for math
\usepackage{subcaption}
\usepackage{epstopdf}
\usepackage{color}
\usepackage{xcolor}
\usepackage[utf8]{inputenc}
\usepackage{amsmath}
\usepackage{booktabs}
\usepackage{multirow}
\usepackage{float}
\usepackage{tabularx}
\allowdisplaybreaks[4]
\usepackage{accents}
\newlength{\dhatheight}

\usepackage{hyperref}
\usepackage{pifont}
\graphicspath{{Figs/}}

\def\figureautorefname~#1\null{Fig.\,#1\null}
\def\tableautorefname~#1\null{Tab.\,#1\null}

\def\equationautorefname~#1\null{Eq.\,(#1)\null}

\begin{document}

\title{Deep Learning to Improve the Sensitivity of Higgs Pair Searches in the $4b$ Channel at the LHC}

\author{Yongcheng Wu}
\email{ycwu@njnu.edu.cn}
\affiliation{Department of Physics, Institute of Theoretical Physics and Institute of Physics Frontiers and Interdisciplinary Sciences, Nanjing Normal University, Nanjing, 210023, China}
\affiliation{Nanjing Key Laboratory of Particle Physics and Astrophysics, Nanjing, 210023, China}
\author{Liang Xiao}
\email{lxiao@njnu.edu.cn}
\affiliation{Department of Physics, Institute of Theoretical Physics and Institute of Physics Frontiers and Interdisciplinary Sciences, Nanjing Normal University, Nanjing, 210023, China}
\author{Yan Zhang}
\email{zyan@njnu.edu.cn }
\affiliation{Department of Physics, Institute of Theoretical Physics and Institute of Physics Frontiers and Interdisciplinary Sciences, Nanjing Normal University, Nanjing, 210023, China}
\affiliation{Nanjing Key Laboratory of Particle Physics and Astrophysics, Nanjing, 210023, China}
% \affiliation{School of Physics Science and Technology, Nanjing Normal University, Nanjing 210023, China}
\preprint{$\begin{gathered}\includegraphics[width=0.05\textwidth]{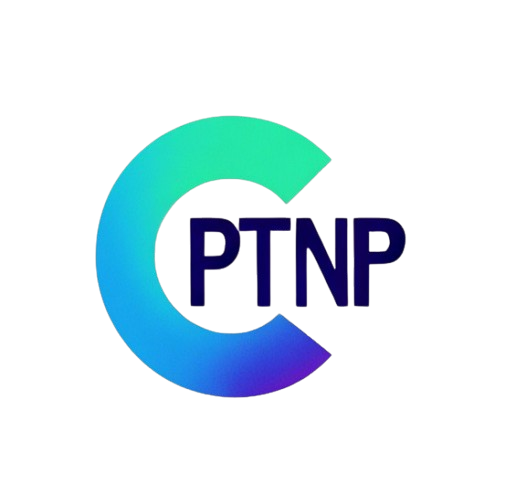}\end{gathered}$\,CPTNP-2025-009}

\begin{abstract}
The Higgs self-coupling is crucial for understanding the structure of the scalar potential and the mechanism of electroweak symmetry breaking. In this work, utilizing deep neural network based on Particle Transformer that relies on attention mechanism, we present a comprehensive analysis of the measurement of the trilinear Higgs self-coupling through the Higgs pair production with subsequent decay into four $b$-quarks ($HH\to b\bar{b}b\bar{b}$) at the LHC. The model processes full event-level information as input, bypassing explicit jet pairing and can serves as an event classifier. At HL-LHC, our approach constrains the $\kappa_\lambda$ to $(-0.53,6.01)$ at 68\% CL achieving over 40\% improvement in precision over conventional cut-based analyses. Comparison against alternative machine learning architectures also shows the outstanding performance of the Transformer-based model, which is mainly due to its ability to capture the correlations in the high-dimensional collision data with the help of attention mechanism. The result highlights the potential of attention-based networks in collider phenomenology.
\end{abstract}

\maketitle
\flushbottom
\clearpage
\tableofcontents
\clearpage
%%%%%%%%%%%%%%%%%%%%%%%%%%%%%%%%%%%%%%%%%%%%%%%%%%
\section{Introduction}% (fold)
\label{sec:introduction}
%%%%%%%%%%%%%%%%%%%%%%%%%%%%%%%%%%%%%%%%%%%%%%%%%%

The Standard Model (SM) is currently the most successful theoretical framework for describing fundamental particles and their interactions, with the Higgs particle being the core of the model. The Higgs field, through the Higgs mechanism, breaks the Electroweak symmetry spontaneously and explains how fundamental particles acquire mass. Since the discovery of the 125 GeV Higgs particle by the ATLAS and CMS experiments in 2012~\cite{ATLAS:2012yve,CMS:2012qbp}, precise measurement of the properties of the Higgs, including the determination of its mass, spin, parity, interactions with other particles and Higgs self-coupling, is one of the most important tasks in particle physics and provides an essential test of the SM. Despite the great success of the SM, there are still phenomena that cannot be explained in SM, such as the existence of dark matter, the nonzero mass of neutrinos, and the matter-antimatter asymmetry etc. These phenomena suggest the existence of the physics beyond the SM (BSM), and the discovery of the Higgs particle has opened a window for exploring new physics. Any deviation between the measured properties of the Higgs particle and the predictions of the SM would be crucial for us to explore the physics beyond the SM.

Currently, the ATLAS and CMS experiments have already performed various measurements about the Higgs. The mass of the Higgs has been precisely measured by both experiments~\cite{CMS:2020xrn,ATLAS:2023oaq}. The width of the Higgs has also been precisely measured through the off-shell effect~\cite{CMS:2022ley,ATLAS:2022vkf} which was originally thought not possible at hadron collider. All major production modes and decay channels of the Higgs have been measured~\cite{CMS:2022dwd,ATLAS:2022vkf} with the overall signal strength in agreement with the SM predictions. From these measurements, the interactions of the Higgs with other SM particles can be extracted which are crucial to test the Higgs mechanism.

Measurement of the Higgs self-coupling~\cite{Abouabid:2024gms,Lu:2015jza,Kling:2016lay,DiMicco:2019ngk,Baur:2003gp,Degrassi:2021uik,Bahl:2023lck,Arco:2022lai,Papaefstathiou:2012qe,Baur:2002rb} is also an important task in probing the properties of the Higgs boson. The Higgs self-coupling determines the shape of the Higgs potential, which is crucial for electroweak symmetry breaking~\cite{Baglio:2012np,Dolan:2012ac}. In the SM, the Higgs potential is expressed as:
 \begin{align}
 V &= \mu^2 \Phi^\dagger \Phi + \lambda (\Phi^\dagger \Phi)^2,
\end{align}
with
\begin{align}
    \Phi = \begin{pmatrix}
       \phi^+\\
       \frac{v+H+i\phi^0}{\sqrt{2}}
    \end{pmatrix}.
\end{align}
where $v=\sqrt{-\mu^2/\lambda}$ is the vacuum expectation value (vev) determined from the potential. The potential for $H$ after the electroweak symmetry breaking can be expressed as
\begin{align}
    V &= \frac{1}{2}m_H^2H^2 + \lambda_{HHH}v H^3 + \frac{1}{4} \lambda_{HHHH}H^4,
\end{align}
where $\lambda_{HHH}=\lambda_{HHHH}=\frac{m_H^2}{2v^2}$ represent the trilinear and quartic Higgs self-couplings. It is clear that in the SM, the Higgs self-couplings are fully determined by the mass and vev of the Higgs which have been measured with a high precision. Any deviation from the SM prediction indicates the presence of new physics~\cite{Durieux:2022hbu,Cheung:2020xij,Chang:2019vez,Gupta:2013zza,Bhattiprolu:2024tsq,Dawson:2015oha,Banerjee:2016nzb}. In this work, we use the kappa framework to parameterize the deviation
\begin{align}
    V &= \frac{1}{2}m_H^2H^2 +\kappa_\lambda \lambda_{HHH}vH^3 + \frac{1}{4}\kappa_{\lambda,4}\lambda_{HHHH}H^4.
\end{align}
where $\kappa_\lambda$ and $\kappa_{\lambda,4}$ indicate the deviations in trilinear and quartic Higgs self-couplings respectively. The Effective Field Theory (EFT) framework provides another powerful and model-independent approach to studying the Higgs self-couplings~\cite{Alasfar:2023xpc,Goertz:2014qta,Azatov:2015oxa,Cao:2015oaa,Li:2019uyy}.

While this study primarily focuses on the trilinear coupling, the quartic Higgs self-coupling is also important. However, direct measurement of this coupling is extremely challenging due to the extremely small cross section of triple Higgs production. Consequently, the measurement of the quartic Higgs self-coupling is difficult at current or near-future colliders such as the LHC or the ILC~\cite{Liu:2018peg}. Future 100 TeV hadron colliders~\cite{Bizon:2018syu,Chen:2015gva,Fuks:2017zkg} or high-energy muon colliders~\cite{Chiesa:2020awd} are considered promising platforms for probing this coupling. Additionally, loop corrections in certain Higgs pair production processes, especially in VBF and VHH channels, can also offer indirect sensitivity to the quartic coupling~\cite{Liu:2018peg}.

Direct measurement of the trilinear Higgs self-coupling relies on the Higgs pair productions~\cite{Torndal:2023mmr,Torndal:2023fky,Davies:2024kvt,Zhang:2024rix,Heinrich:2024dnz,Carvalho:2015ttv,Kim:2018cxf,Bizon:2024juq,Nakamura:2017irk} to which the gluon-gluon fusion (ggF) process provides the dominant contribution at the LHC~\cite{Chen:2019lzz,Baglio:2020ini,Baglio:2018lrj,Borowka:2016ehy}. The Higgs pair production from vector boson fusion (VBF) has also been investigated for the measurement of Higgs self-couplings~\cite{Baglio:2012np,Czurylo:2023nxf}.
In addition to ggF and VBF production of a pair of Higgs, many other processes are also considered to study the Higgs self-coupling including the double Higgs-strahlung process~\cite{Cao:2015oxx}, Higgs pair associated with two top quarks~\cite{Baglio:2012np}, Higgs pair plus jets production~\cite{Chai:2022zeq,Dolan:2013rja}. These processes have smaller cross sections than ggF production, but they can still contribute to the Higgs self-coupling measurement at future colliders with higher energy and luminosity.
On the other hand, the Higgs pair production can be significantly altered in extended Higgs sectors where the Higgs pair is produced resonantly~\cite{Abouabid:2021yvw,Godunov:2014waa,Kotwal:2015rba,DiLuzio:2017tfn,Grober:2017gut}. The studies of the Higgs pair production can also help probing the scalar potential in these models, e.g. xSM~\cite{Huang:2017jws,No:2013wsa,Barger:2014taa,Chen:2014ask}, 2HDM~\cite{Barducci:2019xkq,Heinemeyer:2024hxa,Arhrib:2024hed,Baglio:2014nea,Hespel:2014sla,Bauer:2017cov} and MSSM~\cite{Arganda:2017wjh}. However, in this work, we will focus on the non-resonant production of Higgs pair and leave the resonant production to future work.
The Higgs self-coupling can also be measured indirectly through its contribution to Higgs production and decay from loop corrections~\cite{Chen:2021pqi,Henning:2018kys,McCullough:2013rea,Gorbahn:2016uoy,Degrassi:2016wml,Gao:2023bll,Degrassi:2017ucl,Alasfar:2022zyr}.

Both ATLAS and CMS have conducted studies on the Higgs pair production process. In the ATLAS analysis, the decay channels $HH \rightarrow b\bar{b}b\bar{b}$~\cite{ATLAS:2023qzf}, $HH \rightarrow b\bar{b}\tau^+\tau^-$~\cite{ATLAS:2022xzm}, and $HH \rightarrow b\bar{b}\gamma\gamma$~\cite{ATLAS:2021ifb} were examined. The results indicate that the $\kappa_{\lambda}$ is constrained to the range $(-0.6, 6.6)$ after combining all these channels~\cite{Zabinski:2023jhr}.
In the CMS analysis, the decay channels $HH \rightarrow b\bar{b}ZZ^*$~\cite{CMS:2022omp}, $HH \rightarrow \text{Multilepton}$, $HH \rightarrow b\bar{b}\gamma\gamma$~\cite{CMS:2020tkr}, $HH \rightarrow b\bar{b}\tau^+\tau^-$~\cite{CMS:2022hgz} and $HH \rightarrow b\bar{b}b\bar{b}$~\cite{CMS:2022cpr} were analyzed. The combined results show that the $\kappa_{\lambda}$ value is constrained to $(-1.24, 6.49)$~\cite{CMS:2022dwd}.
Future colliders are expected to significantly improve the measurement precision under higher center-of-mass energy and integrated luminosity. The HL-LHC~\cite{Chang:2018uwu,ATLAS:2025wdq,He:2015spf,Barr:2013tda} is expected to have a precise measurement of the Higgs self-coupling $\kappa_{\lambda} = 1.0^{+0.48}_{-0.42}$, while the HE-LHC~\cite{Goncalves:2018qas}
and FCC-hh~\cite{Taliercio:2022maa,Braibant:2022ebu,Mangano:2020sao,Stapf:2023ndn} are projected to achieve a precision of 5\% on the Higgs self-coupling. The future lepton collider can also probe the Higgs self-coupling with about 20\% precision from FCC-ee~\cite{Li:2022zyn,Li:2024mrd}, ILC~\cite{Kawada:2022jac,List:2024ukv} as well as multi-TeV muon collider~\cite{Buonincontri:2022ylv,Buonincontri:2021okq,Han:2020pif,Buonincontri:2024tpa}.

The Higgs boson predominantly decays into $b\bar{b}$, with a branching ratio of approximately 58\%. Among all possible decay channels, the $4b$ final state constitutes the largest fraction (about 33.6\%) for a Higgs pair system~\cite{Goertz:2013kp}, making it a promising channel for investigation. The current constraints on the Higgs self-coupling are relatively broad from $4b$ channel $\kappa_\lambda\in(-3.3, 11.4)$~\cite{Zabinski:2023jhr}, which is weak compared with other channels and mainly limited by the complex background at the LHC.
Improving the discriminating power between signal and background in $4b$ channel will hence significantly enhance the sensitivity. There are already many studies focusing on improving the sensitivity in this channel~\cite{Behr:2015oqq,Wardrope:2014kya,FerreiradeLima:2014qkf,Li:2019tfd,Alves:2019igs,Amacker:2020bmn,Chiang:2024pho}. In this work, we explore the possibility of using machine learning techniques to tackle the challenge in $4b$ channel of Higgs pair searches at the LHC.

The amount of data generated at colliders is enormous, and machine learning (ML) is particularly well-suited for handling such large-scale datasets~\cite{Alison:2019kud,Andrews:2019faz,Andrews:2018nwy}.
There have been various ML applications in particle physics~\cite{Baldi:2014kfa,Abdughani:2019wuv}, such as jet-tagging~\cite{Guest:2016iqz,Kasieczka:2019dbj,Woodward:2024dxb,Lin:2018cin} including CNN based on image data ~\cite{Cogan:2014oua,deOliveira:2015xxd,Lee:2019cad,Madrazo:2017qgh}, GNN based on graph data~\cite{Dreyer:2020brq,Abdughani:2018wrw,Gong:2022lye}, ParticleNet and Energy Flow Network based on particle clouds~\cite{Qu:2019gqs,Mikuni:2021pou,Komiske:2018cqr,Usman:2024hxz}, and the current state-of-the-art (SOTA) Transformer-based model, Particle Transformer (\textit{ParT})~\cite{Qu:2022mxj,Tagami:2024gtc,Camagni:2024zzi,Wu:2024thh,He:2023cfc}. Similarly, for the measurements of Higgs self-coupling, ML techniques are also used~\cite{Amacker:2020bmn,Chiang:2024pho,Apresyan:2022tqw,Stylianou:2023xit,Alasfar:2022vqw}.
Additionally, there are studies that investigate the performance of different machine learning algorithms in Higgs self-coupling measurement~\cite{Huang:2022rne,Tannenwald:2020mhq,Abdughani:2020xfo}.

For the $HH \rightarrow b\bar{b}b\bar{b}$ channel,
different machine learning architectures have been applied.
For instance, Ref.~\cite{Amacker:2020bmn} employing a DNN architecture demonstrated that, with $\mathcal{L}=3000\,\rm fb^{-1}$ at HL-LHC,
the Higgs self-coupling can be constrained within the range $\kappa_\lambda\in(-0.8, 6.6)$ at 68\% CL. A more recent work~\cite{Chiang:2024pho} utilizing a Transformer model achieved a constraint of \(\kappa_{\lambda} \in (-1.56, 7.57)\) at 95\% CL with $\mathcal{L}=300\,\rm fb^{-1}$ at the LHC.
Considering the excellent performance of \textit{ParT} in jet tagging tasks~\cite{Qu:2022mxj}, in this work, a modified version of \textit{ParT} including entire event information as input for event classification will be used to improve the sensitivity on the Higgs self-coupling through $HH\to b\bar{b}b\bar{b}$ channel.

This paper is organized as follows. In Sec.~\ref{sec:event generation}, we briefly describe all the processes we will consider in this work.
In Sec.~\ref{sec:transformer on event}, after the introduction of the structure of {\it ParT}, we will discuss the modification we will made and the details of the training process. The performance of the modified model on event classification will also be presented. In Sec.\ref{sec: higgs self-coupling constraints}, we present the results on the Higgs self-coupling measurement with a comparison among different methods. The interpretability of the model is also discussed at the end of this section.
Finally, we summarize in Sec.\ref{sec: conclusions}.

\section{Event Generation}
\label{sec:event generation}

\begin{figure}[!tbp]
\centering
\includegraphics[width=0.40\textwidth]{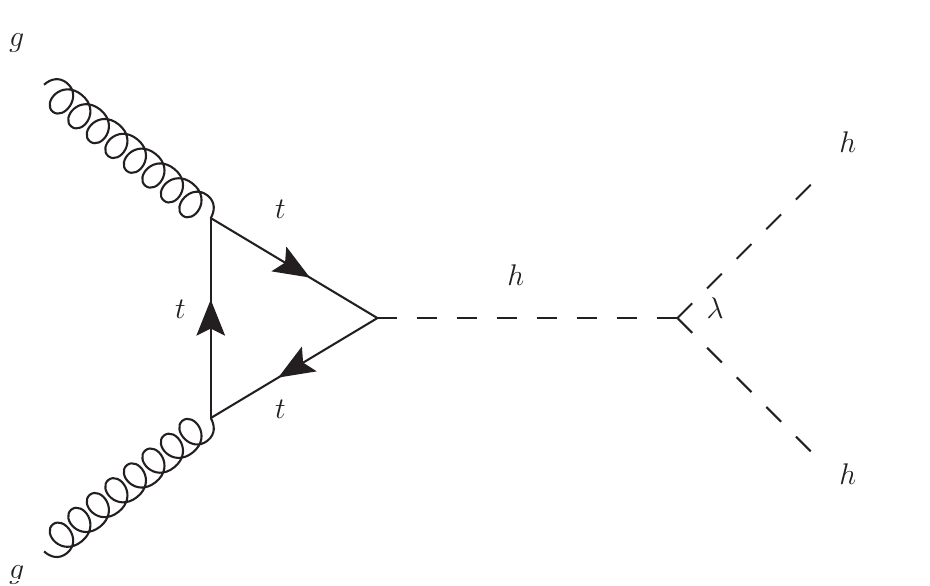}
\includegraphics[width=0.30\textwidth]{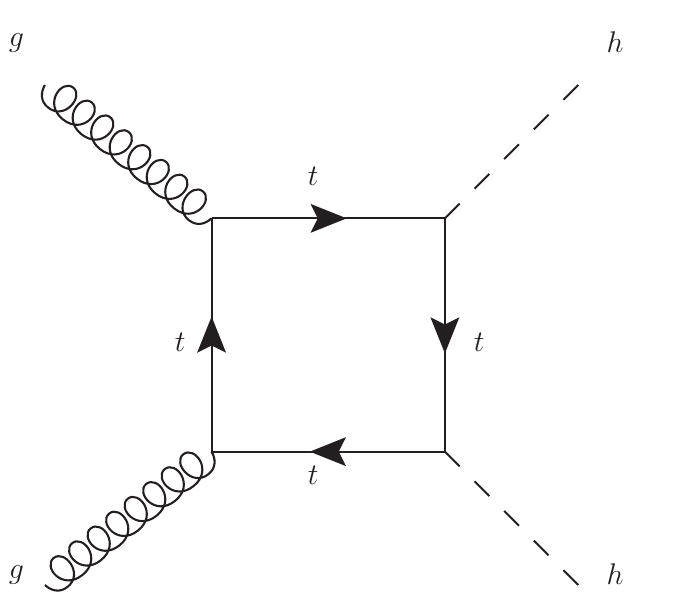}
\caption{The leading-order Feynman diagrams of the di-Higgs production in the SM.}
\label{fig:gg_hh_feynman_diagram}
\end{figure}

\begin{figure}[!b]
    \centering
    \includegraphics[width=0.5\textwidth]{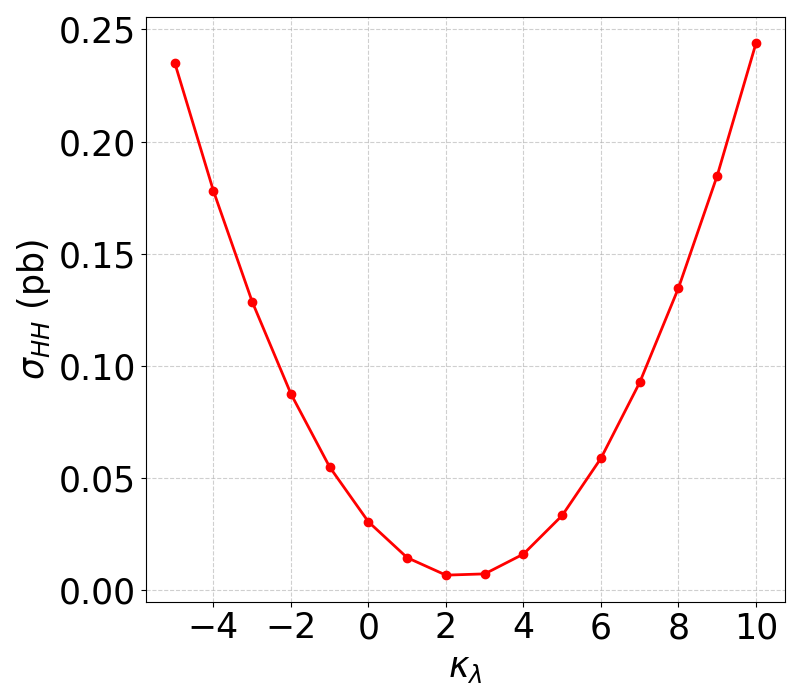}
    \caption{Relationship between the Higgs pair production cross section and $\kappa_\lambda$ at the LHC ($\sqrt{s}=13\,\rm TeV$), assuming all other couplings are fixed.}
    \label{fig:HH_CS_kappa}
\end{figure}

For the Higgs self-coupling measurement, we focus on the non-resonant gluon-gluon fusion (ggF) production of Higgs pair\cite{Kerner:2024dgm,Li:2024iio,Frederix:2014hta}, with both Higgs decays into $b$-quark pair, which contains, at leading order (LO), the contributions from triangle and box diagrams as well as the interference between them. The representative Feynman diagrams are shown in~\autoref{fig:gg_hh_feynman_diagram}. The triangle diagrams depend on the trilinear Higgs self-coupling, while the box diagrams depend on the Yukawa couplings of the fermion in the loop. In our simulations, we include the contributions from the third generation quarks, but with fixed Yukawa coupling at corresponding SM value. On the other hand, the trilinear Higgs self-coupling can differ from the SM value and will influence the total cross section as well as distributions from which we can extract information about the Higgs self-coupling. \autoref{fig:HH_CS_kappa} shows the dependence of the Higgs pair production cross section on $\kappa_\lambda$.
We also parameterize this dependence as follows:
\begin{align}
    \label{equ:cs_hh}
    \sigma_{HH}(\kappa_\lambda) &= (9.96 \times 10^{-3})\,\kappa_{\lambda}^{2}
    - (4.85 \times 10^{-2})\,\kappa_{\lambda}
    + (7.33 \times 10^{-2})\,\rm pb
\end{align}

For the backgrounds, we will mainly focus on the processes containing multi $b$-quarks and/or multi-jets. In our analysis, the main backgrounds are the QCD production of four $b$ quarks ($4b$) as well as two $b$ quarks together with two light quarks ($2b2j$). Other than these two major backgrounds, we also consider the production of $b$-quarks from the decay of $Z$ boson, Higgs, and top quark. In~\autoref{tab:processes_cs}, we list all signal and background processes considered in this analysis together with their cross section at LO. For the signal, as well as $t\bar{t}$, $4b$ and $2b2j$ background processes, the NLO k-factors are also listed~\cite{Amacker:2020bmn,Heinrich:2019bkc,deFlorian:2013jea,deFlorian:2015moa,deFlorian:2013uza}.

\begin{table}[!tbp]
    \centering
    \begin{tabular}{|c|c|c|}
    \hline\hline
    Process & Cross Section [fb] & NLO k-factor \\
    \hline
    $HH$ & $1.454\times 10^1$  & 2.4 \\
    $b\bar{b}b\bar{b}$ & $2.475\times 10^6$ & 1.3 \\
    $b\bar{b}jj$ & $5.670\times 10^8$ & 1.6 \\
    $t\bar{t}$ & $5.058\times 10^5$ & 1.5 \\
    $t\bar{t}b\bar{b}$ & $1.356\times10^4$ & - \\
    $t\bar{t}H$ & $3.998\times 10^2$ & - \\
    $b\bar{b}H$ & $4.729\times 10^1$ & - \\
    $ZZ$ & $9.340\times10^3$ & - \\
    $ZH$ & $5.799\times10^2$ & - \\
    \hline
    \end{tabular}
    \caption{All processes considered in this work, with the LO cross section at $\sqrt{s}=13\,\rm TeV$. For $HH$, $4b$ and $2b2j$ processes, the NLO k-factor are also listed.}
    \label{tab:processes_cs}
\end{table}

The events for both signal process with possible different values of $\kappa_\lambda$ and background processes are simulated by {\tt MadGraph}~\cite{Alwall:2014hca} with $\sqrt{s}=13\,\rm TeV$. To increase the simulation efficiency, we use a moderate parton-level basic cuts: $p_T^{j,b} > 20\,\rm GeV$, $|\eta^{j,b}|<4$. The {\tt Pythia}~\cite{Sjostrand:2014zea} is used for the hadronization and showering followed by the detector simulation by {\tt Delphes}~\cite{deFavereau:2013fsa}. {\tt FastJet}~\cite{Cacciari:2011ma} is further linked to reconstruct the Jets from the particle-flow output from {\tt Delphes} using anti-$k_t$ algorithm~\cite{Cacciari:2008gp} with two different values of $\Delta R$, 0.5 for a slim jet originated from quarks/gluon, 1.0 for a fat jet originated mainly from heavier objects, e.g. Higgs, W/Z boson, top quarks.

\section{Transformer on Event}
\label{sec:transformer on event}

In this studies, instead of working on the individual objects (jets, charged leptons, photons etc.) within an event, we investigate the possibility of directly working on the whole event. The Particle Transformer ({\it ParT})~\cite{Qu:2022mxj} is the state-of-the-arts (SOTA) object tagging machine learning (ML) algorithm, which was trained on a 10-label-classification task. By implementing {\it ParT}, we could in principle tag the Higgs, top, W/Z boson as a fat jet in the events, and then combine them to extract the event of signal (Higgs pair) from other background processes (containing top quark, W/Z boson etc.).
Although, the individual reconstruction efficiency is sufficient, relying on reconstructed individual heavy objects inside one events may suffer from the combinatorial problem which can heavily reduce the sensitivity, especially for $HH\to b\bar{b}b\bar{b}$ case where the four $b$-jets in the final states are indistinguishable and there are 6 combinations of these $b$-jets to form a Higgs pair candidate.
Hence, we explore the possibility of training on the information of entire event which can be treated as a single extremely fat jet.
Consequently, the {\it ParT} framework can still be used with minor modification in the event classification task. In the following, we will first briefly introduce the main features of {\it ParT}. Then the training details in our setup will be presented. The performance on the event classification is also discussed. To avoid confusion with the original {\it ParT} model, we will refer to the {\it ParT} working on the entire event in our setup as Event Transformer ({\it EvenT}).

\subsection{Basics about Particle Transformer}

The {\it ParT}~\cite{Qu:2022mxj} is basically a classification model based on the {\it Transformer} architecture~\cite{vaswani2023attentionneed} which is a deep learning model originally designed for natural language processing (NLP) tasks. The core of {\it Transformer} is the self-attention mechanism, which establishes relationships among all elements in the input sequence, rather than being limited to local information like CNN. By computing the attention weight matrix, {\it Transformer} can capture global information. Due to the advantages of this mechanism, {\it Transformer} has demonstrated powerful performance across multiple domains.

\begin{figure}[!tbp]
    \centering
    \includegraphics[width=\textwidth,trim=0 150 0 50]{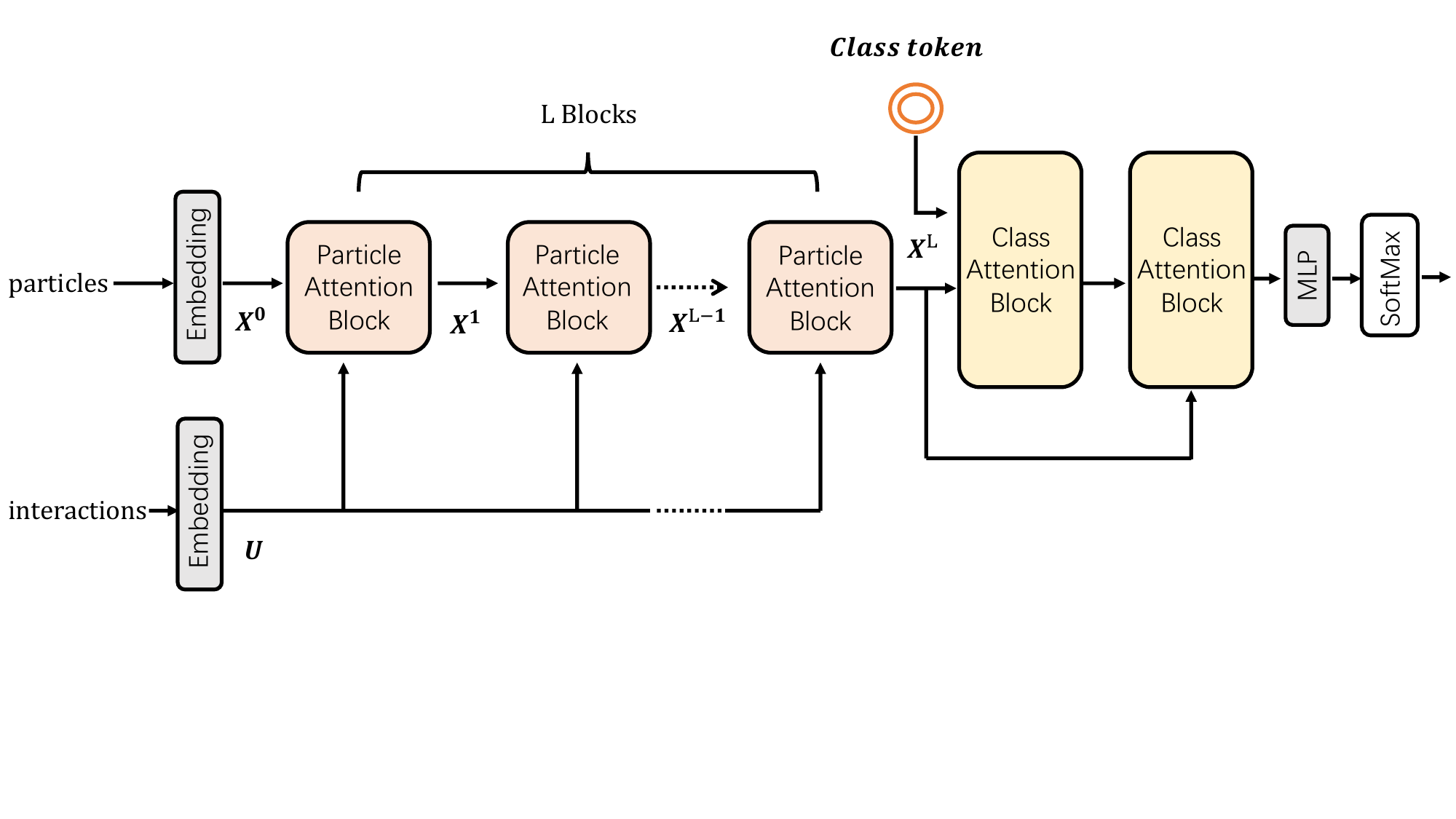}
    \caption{The {\it ParT} Model Architecture.
    }
    \label{fig:ParT_Architecture}
\end{figure}

The structure of {\it ParT} is shown in~\autoref{fig:ParT_Architecture}. The model receives two sets of inputs: {\it particles} which include the features (kinematics, PID, trajectory displacment etc.) of each single object (tracks, charged lepton etc.) and {\it interactions} which indicate the relationships between two objects. In {\it ParT}, there are mainly 4 such relationships:
\begin{align}
\Delta R_{ij} &= \sqrt{(y_i - y_j)^2 + (\phi_i - \phi_j)^2}, \\
k_{T,ij} &= \min(p_{T,i}, p_{T,j}) \Delta R_{ij}, \\
z_{ij} &= \frac{\min(p_{T,i}, p_{T,j})}{p_{T,i} + p_{T,j}}, \\
m_{ij}^2 &= (p_i+p_j)^2.
\end{align}
where $y,\phi$ are the rapidity and azimuthal angle of the individual object respectively. $p_T$ is the corresponding magnitude of the transverse momentum. Then the inputs are passed through the intermediate layers consisting of {\it Particle Attention Blocks} and {\it Class Attention Blocks}. Compared to the traditional {\it Transformer}, {\it ParT} introduces, in the {\it Particle Attention Block}, new attention mechanism referred as particle multihead attention (P-MHA) where the {\it interactions} are included in the attention calculation:
\begin{align}
\text{P-MHA}(X) &= \text{concat}(H_1, \dots, H_h) W^O, \\
H_i &= \mathcal{A}_iV_i, \label{eq:pmha}\\
\mathcal{A}_i &= \text{softmax}\left( \frac{Q_i K_i^\top}{\sqrt{d_k}} + U_i \right), \label{equ:attention_matrix}\\
Q_i &= X W_i^Q+b^{Q}_{i}, \\
K_i &= X W_i^K+b^{K}_{i}, \\
V_i &= X W_i^V+b^{V}_{i}.
\end{align}
where $X$ includes the {\it particles} features, $U$ contains features of the {\it interactions} between objects. $W_i^j$ and $b_i^j$ are learnable parameters. Including the {\it interactions} $U$ in the attention calculation enables the model to better capture the relationships between particles and enhance the expressiveness of the attention mechanism. $\mathcal{A}_i$ is the attention weight matrix of which each element represents the attention score of each pair of input {\it particles}.

The {\it Class Attention Block} shares a similar architecture with the Particle Attention Block but differs in two main aspects: it uses the standard Multi-Head Attention (MHA) mechanism instead of the particle-specific variant, and the MHA takes a global class token as input along with the particle embeddings. The query, key, and value matrices are then computed as:
\begin{align}
Q &= W_q X_{\text{class}} + b_q, \\
K &= W_k \mathbf{Z} + b_k, \\
V &= W_v \mathbf{Z} + b_v,
\end{align}
where $\mathbf{Z} = [X_{\text{class}}, X_{\rm PAB}]$ is the concatenation of the class token and the particle embeddings output from the last Particle Attention Block. The output of the {\it Class Attention Block} are then passed to a Softmax layer which produces the probabilities over the classes.

\subsection{Training Details}

In our setup, the {\it EvenT} model receive the information from the whole event as input features. To further facilitate establishing connections among objects within an event, information from jets reconstructed using two different $\Delta R = 0.5$ and $1.0$, as discussed in last section, is also provided.
The {\it EvenT} model is trained from scratch on a dataset with 9 classes listed in~\autoref{tab:processes_cs} containing 3 million samples per class. After each training epoch, the model is validated on a separate validation set containing 3 million samples per class, while its final performance is evaluated on an independent test set with an equivalent sample size per class.
The model is trained with a batch size of 128 and initial learning rate of 0.001.
The learning rate remains constant during the first 70\% of training epochs, after which it decays exponentially, reaching 1\% of its initial value by the end of training.
The number of training epochs is set to 50, and the number of heads in the multi-head attention mechanism is 8.

In general, all processes during the training shall be treated equally. However, for our purpose, we would like to single out the Higgs pair process as much as possible from the other processes among which $2b2j$ process has the largest cross section. In order to avoid large contamination from $2b2j$ process,
unlike \textit{ParT} which is used for a general purpose jet-tagging task, we adopt a weighted cross-entropy loss function, where the process $2b2j$ is assigned with a weight of 70, while all other processes are assigned with a weight of 1.
After applying the weighted cross-entropy loss, the probability of miss-classifying $2b2j$ events as $HH$ events is reduced from $5.17\times10^{-3}$ to $5.26\times10^{-5}$ by two orders of magnitude which dramatically improves the analysis about the Higgs pair process. The measurement uncertainty of the Higgs pair production cross section can be reduced from about 400\% to around 200\% by such improvement.

\begin{figure}[!tbp]
    \centering
    \includegraphics[width=0.9\textwidth]{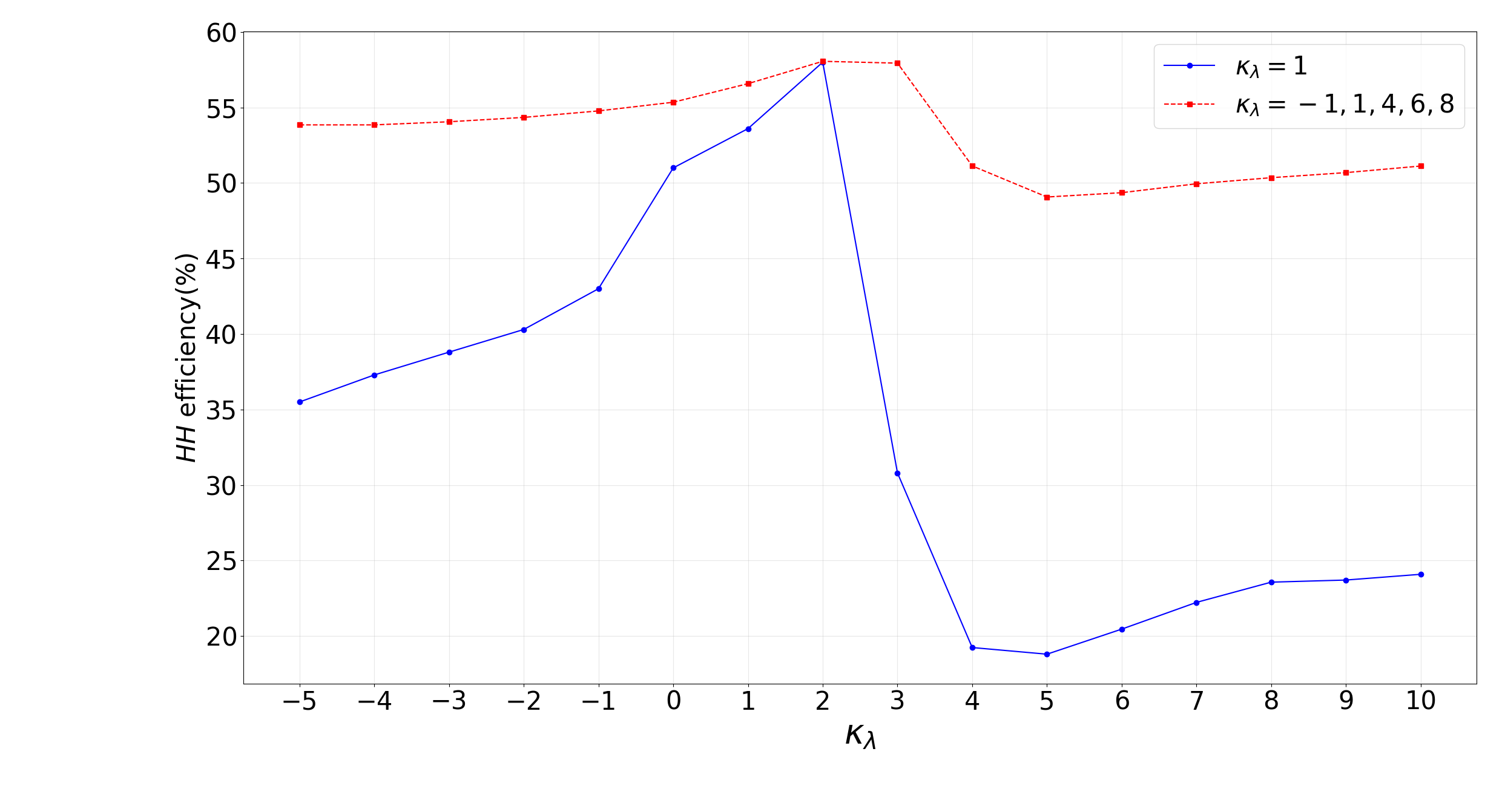}
    \caption{$HH$ efficiency under different $\kappa_{\lambda}$ values for training containing only $\kappa_\lambda=1$ $HH$ data (blue line) and training containing $\kappa_\lambda=-1,1,4,6,8$ $HH$ data (red line). }
    \label{fig:signal-efficiency-differ-kappa}
\end{figure}

Further, as our target is the measurement of $\kappa_\lambda$, the model is required to work properly on all different value of $\kappa_\lambda$ in order to obtain the optimal result. As a comparison, we show in~\autoref{fig:signal-efficiency-differ-kappa} the results of $HH$ efficiency from training on data set with only $\kappa_\lambda = 1$ (blue line) and training on data set containing $\kappa_\lambda = \{-1,1,4,6,8\}$ (red line). It is clear that for the training using data with only $\kappa_\lambda = 1$, the $HH$ efficiency will dramatically decrease for other value of $\kappa_\lambda$. The variance of the efficiency can be as large as 40\% (from 58\% at $\kappa_\lambda=2$ to 18\% at $\kappa_\lambda = 3$). On the other hand, when we include the data with other value of $\kappa_\lambda$, the $HH$ efficiency is roughly stable at around 50\%. Such performance is hence suitable for the further analysis on the measurement precision about $\kappa_\lambda$.

\subsection{Performance of Event Classification}
\label{sec:performance}

In this section, we further demonstrate the classification performance of the model.
{\it EvenT} is now essentially a multi-class classification model with 9 event categories.
For multi-class models, the performance can be evaluated using the confusion matrix, ROC curves and AUC scores.

\begin{figure}[!tb]
    \centering
        \includegraphics[width=0.48\textwidth]{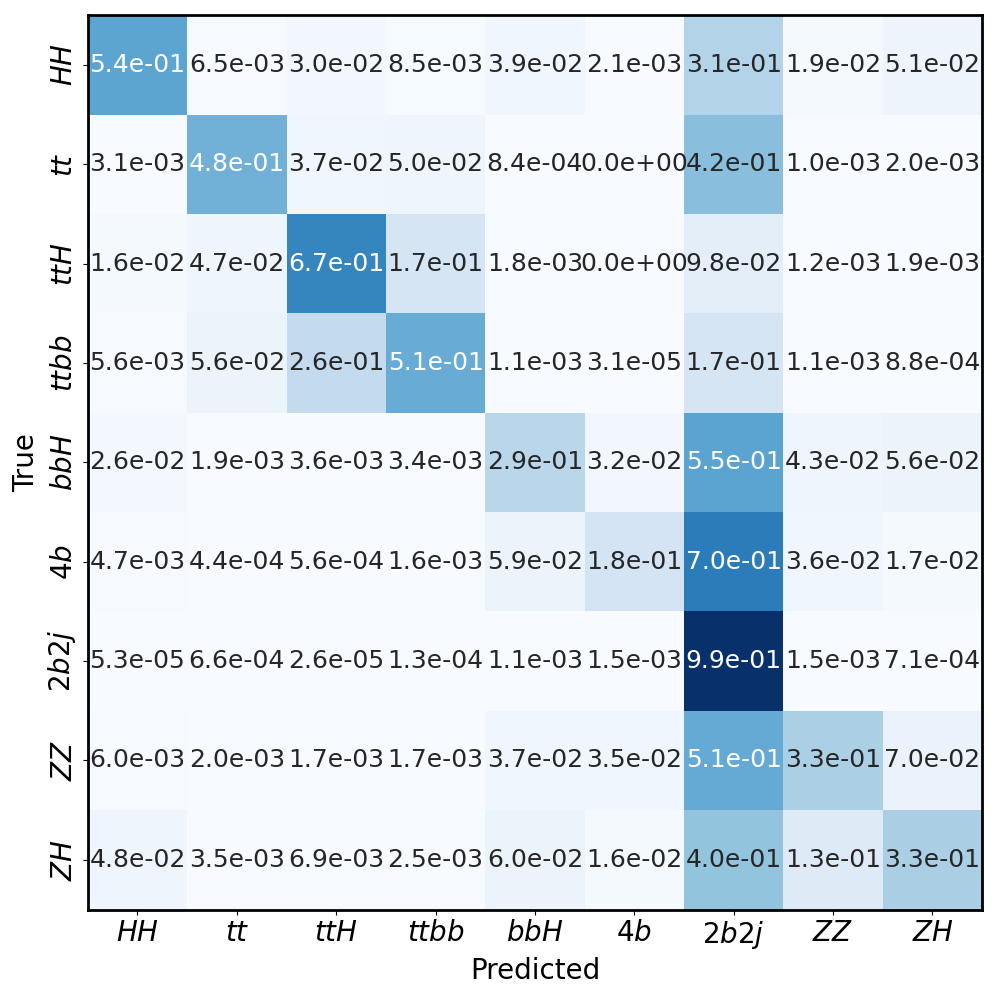}
        \includegraphics[width=0.49\textwidth]{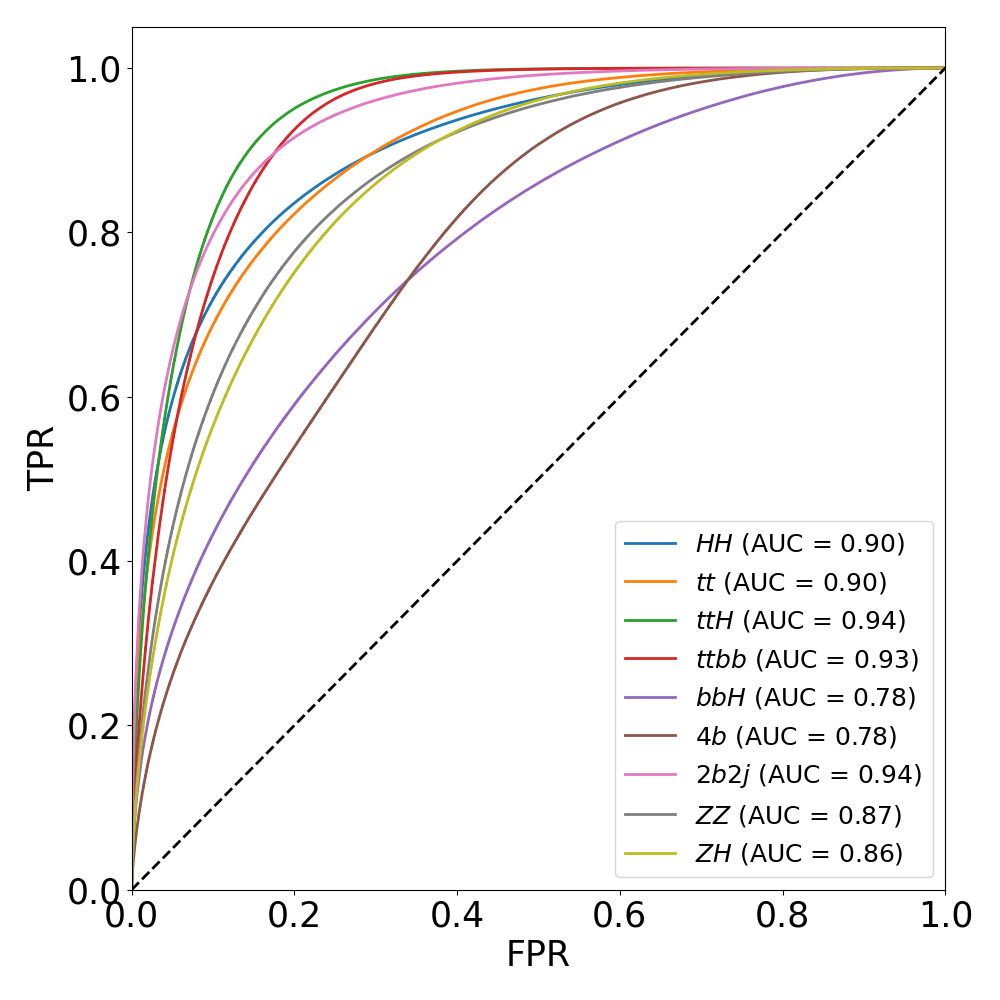}
    \caption{Left: Confusion matrix of the {\it EvenT} on 9-class classification task. Right: OvR ROC curves for each class of {\it EvenT} and the corresponding AUC. All the results are obtained with the weighted cross-entropy loss.}
    \label{fig:CM-ROC}
\end{figure}

A confusion matrix is structured as an $N \times N$ matrix, where N denotes the number of target classes, with rows representing true labels and columns indicating predicted labels, to quantify the performance of the classification model. The confusion matrix can be presented as:
\begin{equation}
\text{Confusion Matrix} =
\begin{bmatrix}
C_{11} & C_{12} & \cdots & C_{1N} \\
C_{21} & C_{22} & \cdots & C_{2N} \\
\vdots & \vdots & \ddots & \vdots \\
C_{N1} & C_{N2} & \cdots & C_{NN}
\end{bmatrix}
\label{equ:confusion_matrix}
\end{equation}
Specifically:
\begin{itemize}
    \item $C_{ii}$ represents the probability that an event of true class $i$ is correctly predicted as class $i$.
    \item $C_{ij}$ (for $i \neq j$) represents the probability that an event of true class $i$ is incorrectly predicted as class $j$.
\end{itemize}
The confusion matrix of {\it EvenT} on the test set, where the $HH$ class only includes events with $\kappa_{\lambda}=1$, is visualized in the left panel of~\autoref{fig:CM-ROC} where the color intensity indicates the corresponding probabilities.

There are several comments about the confusion matrix in order. First, it is clear from the $2b2j$ column, every class has a relatively high efficiency tagging as $2b2j$ since we enhance the weight of $2b2j$ events during the training as discussed above. Second, as can be seen from the diagonal elements, when not considering the $2b2j$ class, the rest classes all have a good self-tagging efficiency. Further, processes similar to each other will form a diagonal block in the tagging efficiency, e.g. $(ttbb, ttH)$ and $(ZZ, ZH)$. For our purpose of measuring Higgs pair production process, the first column of the confusion matrix indicates that all other processes have been heavily suppressed while keep a sufficient efficiency of $HH$ process.

The ROC (Receiver Operating Characteristic) curve and AUC (Area Under Curve) are commonly used metrics for binary classification tasks. They can also be extended to multi-class classification problems for model evaluation.
In a binary classification setting with only positive and negative classes, the ROC curve is generated from the True Positive Rate (TPR) and the False Positive Rate (FPR) at various classification thresholds.
The AUC is the area under the ROC curve and is used to quantify the overall classification performance.

For multi-class classification tasks, the One-vs-Rest (OvR) strategy can be used to simplify the problem into a series of binary classification tasks, allowing the computation of ROC curves and AUC scores. For each class, it is treated as the positive class while all other classes are treated as the negative class.
The right panel of \autoref{fig:CM-ROC} shows the ROC curves and the corresponding AUC values for each class. For most of the classes, the AUC is around 0.9 indicating a relatively good performance of the {\it EvenT}. Further, the AUC for distinguishing $HH$ process from other processes is about $0.9$, this provides a solid foundation for the analysis of the Higgs self-coupling via Higgs pair production in the next section.

\section{Higgs Self-Coupling Measurement}
\label{sec: higgs self-coupling constraints}

\subsection{{\it EvenT}-Based Analysis}
\begin{figure}[!tbp]
    \centering
    \includegraphics[width=0.9\textwidth]{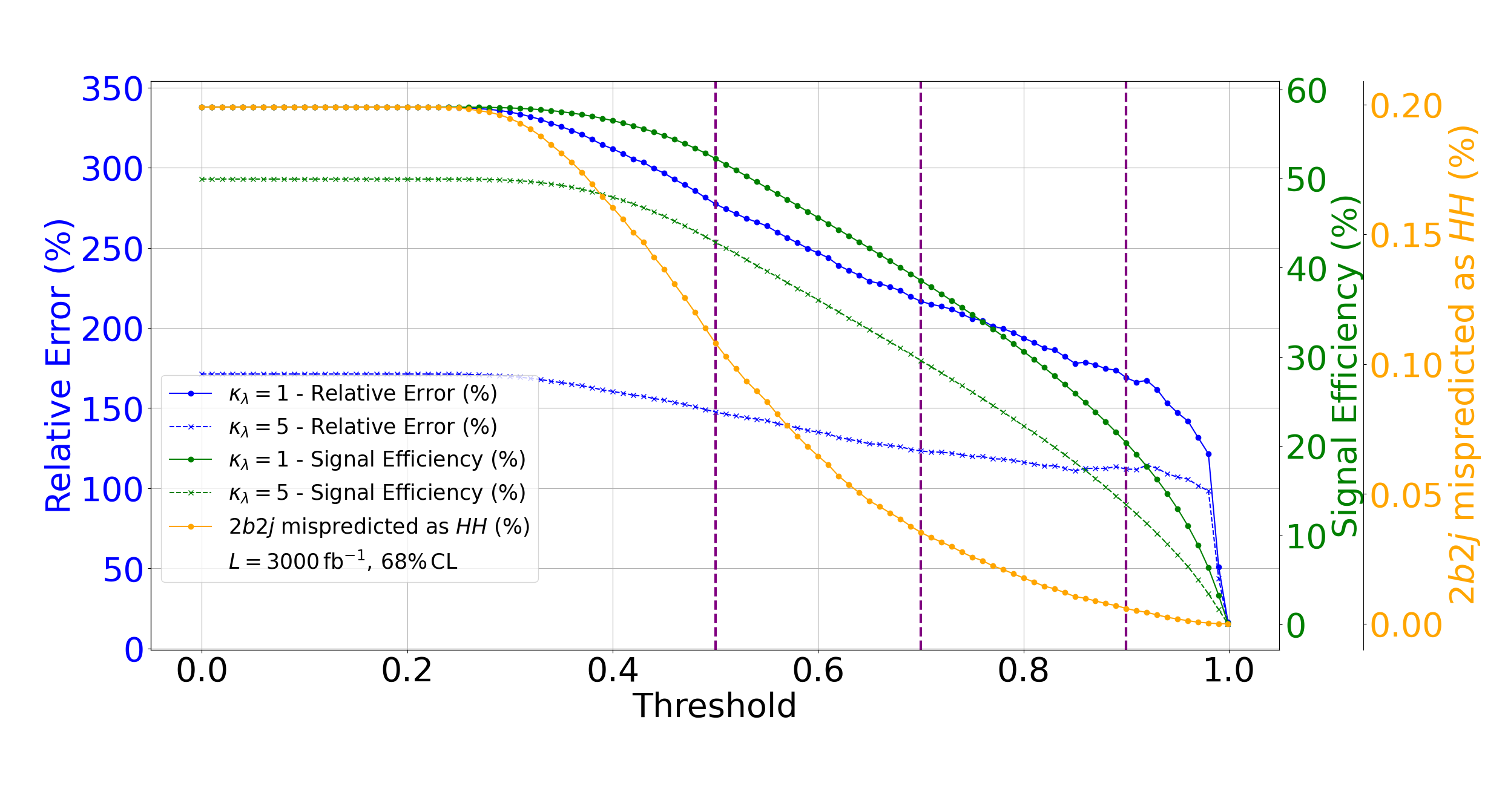}
    \caption{Signal efficiency (green lines) and relative error of the Higgs pair production cross-section measurement (blue lines) for $\kappa_\lambda = 1$ (solid lines) and $\kappa_\lambda = 5$ (dashed lines) at different thresholds. The misclassification rate of the dominant background $2b2j$ is also shown in orange.}
    \label{fig:acc-error-vs-threshold}
\end{figure}

Based on the {\it EvenT} discussed in the above section, we will consider the measurement of $\kappa_\lambda$ through the $HH\to 4b$ channel in this section.
The corresponding $\chi^2$, which indicates the deviation from the SM case, is constructed according to
\begin{align}
    \label{equ:chi2}
    \chi^2(\sigma_{HH},\kappa_\lambda) &= \frac{(S(\sigma_{HH},\kappa_\lambda)-S_{SM})^2}{S_{SM}}\\
    S(\sigma_{HH},\kappa_\lambda) &= \mathcal{L}\times\left(C_{11}(\kappa_\lambda)\sigma_{HH} +\sum_{i=2}^{9} C_{i1}\sigma_i\right)\\
    S_{SM} &= S(\sigma_{HH}|_{\kappa_\lambda=1},\kappa_\lambda = 1)
\end{align}
where $i=1,\cdots,9$ corresponds to all the processes considered in the confusion matrix shown in the left panel of~\autoref{fig:CM-ROC}, including: $HH$, $tt$, $ttH$, $ttbb$, $Hbb$, $4b$, $2b2j$, $ZZ$, and $ZH$.
$C_{ij}$ represents the elements in the confusion matrix. For $HH$ process, the dependence of $C_{11}$ on $\kappa_\lambda$ is also included as shown in~\autoref{fig:signal-efficiency-differ-kappa}.
$\sigma_i$ denotes the cross section of each process listed in~\autoref{tab:processes_cs}. Note that $\sigma_{HH}$ will be a free parameter in the $\chi^2$ calculation.
In our analysis, we consider the LHC experiment with $\sqrt{s} = 13\,\rm TeV$ and integrated luminosities of $\mathcal{L} = 300/3000\,\rm fb^{-1}$.
Based on the above setup, for a fixed value of $\kappa_\lambda$, the $\chi^2$ depends on the corresponding Higgs pair production cross section $\sigma_{HH}$. Then, the upper limit on the cross section $\sigma_{HH}$ can be obtained for given $\kappa_\lambda$.
However, as we discussed in above section, the performance of {\it EvenT} also depends on the classification threshold. The effect of the threshold on the Higgs pair production measurement is shown in~\autoref{fig:acc-error-vs-threshold}. We can see that the signal efficiency decrease smoothly with the increase of the threshold for both $\kappa_\lambda=1$ (solid green line) and $\kappa_\lambda=5$ (dashed green line). While the misclassification rate of the dominant background $2b2j$ (solid orange line) drops more sharply. Hence, the relative error of measuring the Higgs pair production cross section will decrease with higher threshold as shown by the blue curves in~\autoref{fig:acc-error-vs-threshold}. However, we need to emphasize that when the threshold is close to 1, the events passing the threshold will decrease dramatically. The corresponding analysis hence suffers from large uncertainties. In our analysis, we will use three benchmark thresholds $p_{\rm th}=0.5,0.7,0.9$ (indicated by the vertical purple dashed lines in~\autoref{fig:acc-error-vs-threshold}) together with $p_{\rm th}=0.0$ where we rely entirely on the raw output of {\it EvenT} for the classification.

\begin{figure}[!tbp]
    \centering
    \includegraphics[width=0.5\textwidth]{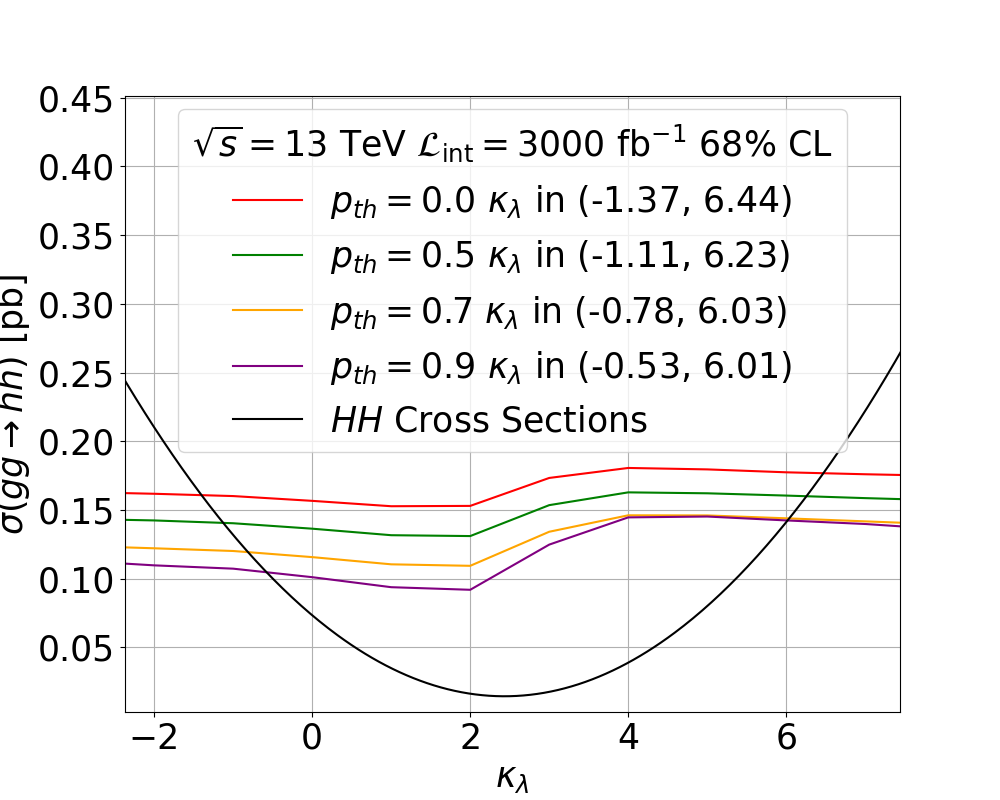}
    \caption{Upper limit on the Higgs pair production cross section as a function of $\kappa_\lambda$ for different classification thresholds $p_{\rm th}$ (four colored lines) with $\sqrt{s}=13\,\rm TeV$ and $\mathcal{L}=3000\,\rm fb^{-1}$ at the LHC. The Higgs pair production cross section as a function of $\kappa_\lambda$ is also shown as black line.}
    \label{fig:diHiggs-comparison}
\end{figure}

The upper limits on the Higgs pair production cross section as a function of $\kappa_\lambda$ obtained from {\it EvenT} with different classification thresholds are shown in~\autoref{fig:diHiggs-comparison}. The upper limits follow well with the $HH$ efficiency shown in~\autoref{fig:signal-efficiency-differ-kappa}. The theoretical prediction of the Higgs pair production cross section is also shown in~\autoref{fig:diHiggs-comparison}. By comparing the {\it EvenT}-derived upper limits with the theoretical prediction, we can extract the constraints on the Higgs self-coupling $\kappa_\lambda$. At the HL-LHC, $\sqrt{s}=13\,\rm TeV$ and $\mathcal{L}=3000\,\rm fb^{-1}$, the constraint at 68\% CL is given as $\kappa_\lambda \in [-0.53, 6.01]$ for $p_{\rm th}=0.9$. The constraints will be a little bit weaker for other choices of the threshold.

\subsection{Cut-Based Analysis}

As a comparison, we followed the cut-based analysis presented in~\cite{ATLAS:2023qzf} to single out $HH\to b\bar{b}b\bar{b}$ events. In this analysis, the events first need to pass several general selections including: requiring at least 4 $b$-tagged (with tagging efficiency about 77\%) jets with $p_T>40\,\rm GeV$, and $|\eta_b|<2.5$. The four $b$-jets with the highest $p_T$ will be used to reconstruct the Higgs pair system. Among the three possible pairings of these four $b$-jets, the one in which the higher $p_T$ jet pair has the smallest $\Delta R$ is used for further analysis. To further suppress the background, $|\Delta\eta_{HH}|<1.5$ is also imposed for the two reconstructed Higgs.

Events are further required to satisfy additional selection criteria designed to reduce the background and improve the analysis sensitivity.
To suppress the $t\bar{t}$ background, a top veto cut is needed. The top veto discriminant $X_{Wt}$ is defined as:
\begin{align}
X_{Wt} = \min_{jjb}\left\{\sqrt{\left(\frac{m_{jj} - m_W}{0.1m_{jj}}\right)^2 + \left(\frac{m_{jjb} - m_t}{0.1m_{jjb}}\right)^2}\right\}
\end{align}
where $m_W = 80.4\,\rm GeV$ and $m_t = 172.5\,\rm GeV$ are the nominal $W$ boson and top quark mass. $m_{jj}$ is the invariant mass of two jets that are assumed to come from the $W$ boson decay. Together with one of the leading $b$-jet, $m_{jjb}$ represent the invariant mass of the reconstructed top. The $X_{Wt}$ is then obtained by minimizing over all the combinations of two normal jets and one $b$-jet. During the minimization, 10\% uncertainties are used to approximate the invariant mass resolution. Then events with $X_{Wt}<1.5$ are excluded in the analysis.

A discriminant $X_{HH}$ is defined to further test the compatibility of events with the $HH\to b\bar{b}b\bar{b}$:
\begin{align}
X_{HH} = \sqrt{\left(\frac{m_{H_1} - 124\,\mathrm{GeV}}{0.1 m_{H_1}}\right)^2 + \left(\frac{m_{H_2} - 117\,\mathrm{GeV}}{0.1 m_{H_2}}\right)^2}
\end{align}
where $m_{H_1}$ and $m_{H_2}$ are the masses of the leading and subleading reconstructed Higgs boson candidates, respectively. The reference mass $124\,\rm GeV$ and $117\,\rm GeV$ are obtained from the $m_{H_1}$ and $m_{H_2}$ distribution from the simulation~\cite{ATLAS:2023qzf}. Note that, when calculating $X_{HH}$, 10\% uncertainties are also used for the reconstructed invariant masses.
Events with $X_{HH} < 1.6$ are considered as $HH\to b\bar{b}b\bar{b}$ signal.

\subsection{Performance}

\begin{figure}[!tbp]
    \centering
    \includegraphics[width=0.8\textwidth]{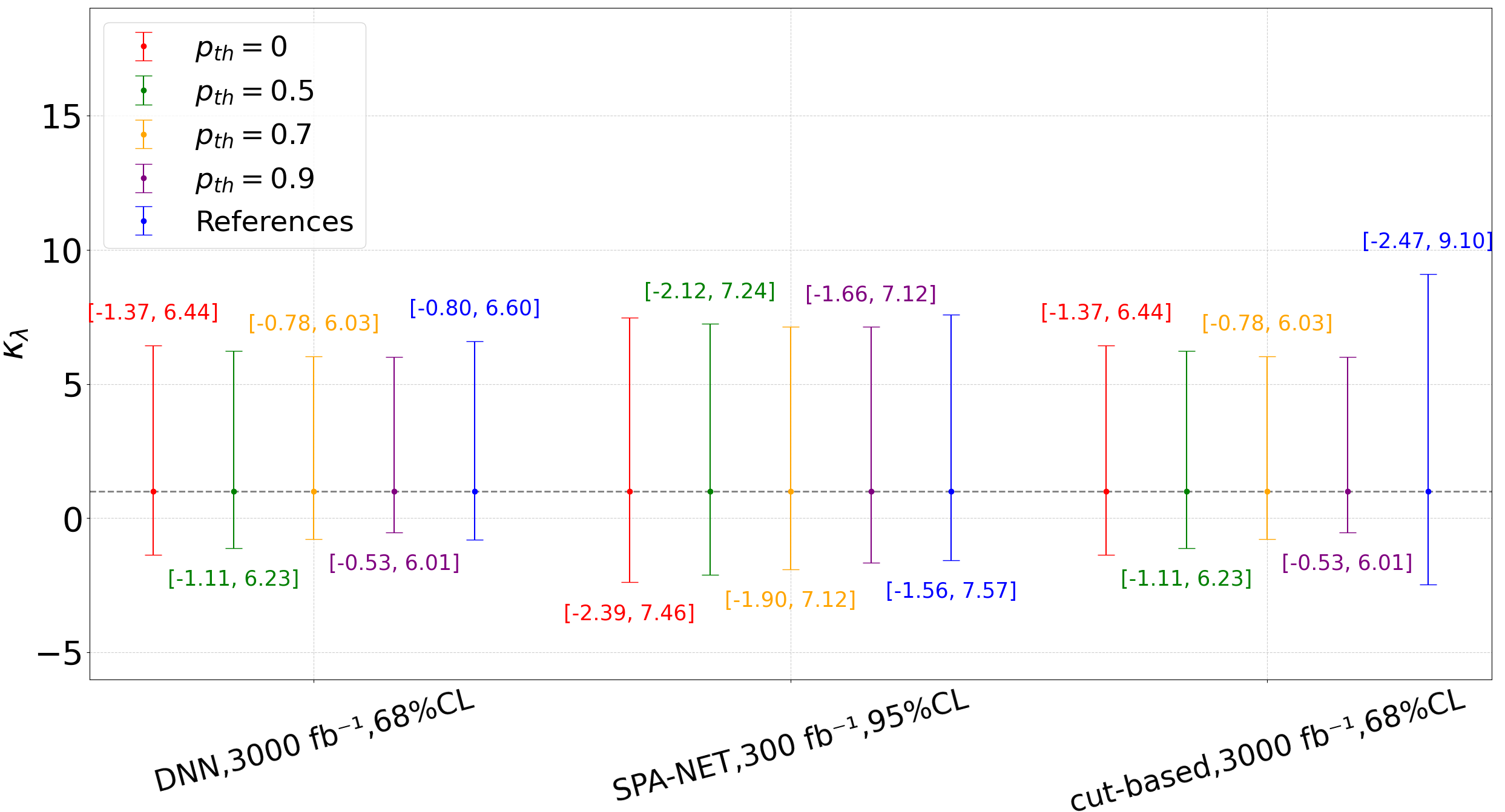}
    \caption{The comparison of the constraints on $\kappa_\lambda$ with three benchmark studies.}
    \label{fig:error_bar}
\end{figure}

To demonstrate the performance of \textit{EvenT}, we compare the results obtained from the \textit{EvenT}-based analysis with those from a cut-based analysis using kinematic variables discussed above, as well as with two ML-based studies, one using DNN~\cite{Amacker:2020bmn}, the other using SPA-NET~\cite{Chiang:2024pho}. The comparison are shown in~\autoref{fig:error_bar}.

The studies in~\cite{Amacker:2020bmn} employs a DNN-based model to single out the Higgs pair signals where they also include comprehensive background studies. The DNN model contains two fully connected hidden layers with 250 hidden nodes with {\tt ReLU} activation function. To prevent overfitting, {\tt Dropout} layer is added between the two hidden layers which randomly drop 30\% of the nodes during training.
The results are given at 68\% CL with an integrated luminosity of $3000\,\rm fb^{-1}$ at the HL-LHC.
The comparison is shown in the left group of~\autoref{fig:error_bar}. It is clear that when $\kappa_\lambda > 0$, even we do not deal with the threshold to enhance the sensitivity, the constraint from {\it EvenT} is stronger than that from the DNN method. While for the negative side, as we only include one negative value in the training $\kappa_\lambda = -1$, without imposing strong threshold, {\it EvenT} is not as good as that from the DNN method. However, when we include a strong threshold to enhance the sensitivity, the result becomes better than that of DNN.

SPA-NET is also an attention-based model~\cite{Fenton:2020woz,Shmakov:2021qdz,Fenton:2023ikr} that employs a stack of transformer encoders to embed input features. These embeddings are then used for jet assignment and event classification via a symmetric tensor attention module, which preserves the permutation symmetry inherent to the input data.
The SPA-NET has also been used in the $HH\to b\bar{b}b\bar{b}$ analysis in~\cite{Chiang:2024pho} where the architecture is used both in pairing of the $b$-jets into two Higgs as well as in signal-background discrimination. However, the analysis in~\cite{Chiang:2024pho} considered only the $4b$ background. Hence, the comparison is made by including only $4b$ background in {\it EvenT} analysis. The results are shown in the middle group of~\autoref{fig:error_bar} which are given at 95\% CL with $300\,\rm fb^{-1}$ luminosity at the LHC. The constraint on the positive side of $\kappa_\lambda$ from {\it EvenT} is slightly better than that from SPA-NET~\cite{Chiang:2024pho}. However, in the negative side, the {\it EvenT} is worse no matter how we choose the threshold. One major reason is that {\it EvenT} is trained with a focus on the dominant $2b2j$ background. Consequently, with only $4b$ background, the result is not fully optimized for {\it EvenT}. Nevertheless, the results from {\it EvenT} and SPA-NET are comparable.

The comparison with the cut-based analysis discussed above is also presented in the right group of~\autoref{fig:error_bar}. Note that the cut-based analysis and {\it EvenT}-based analysis are performed on the same set of testing data. The constraints on $\kappa_\lambda$ are given at 68\% CL with $3000\,\rm fb^{-1}$ at 13 TeV. It is clear that, the {\it EvenT} result is already better than that from the cut-based analysis without imposing any threshold to enhance the sensitivity. With stronger threshold in the classification, the result from {\it EvenT} becomes much stronger than that from the general cut-based analysis which clearly demonstrates the advantage of the ML models in the analysis.

\subsection{Model Interpretability}

\begin{figure}[!tbp]
    \centering
    \begin{minipage}[b]{0.48\textwidth}
        \centering
        \includegraphics[width=\textwidth]{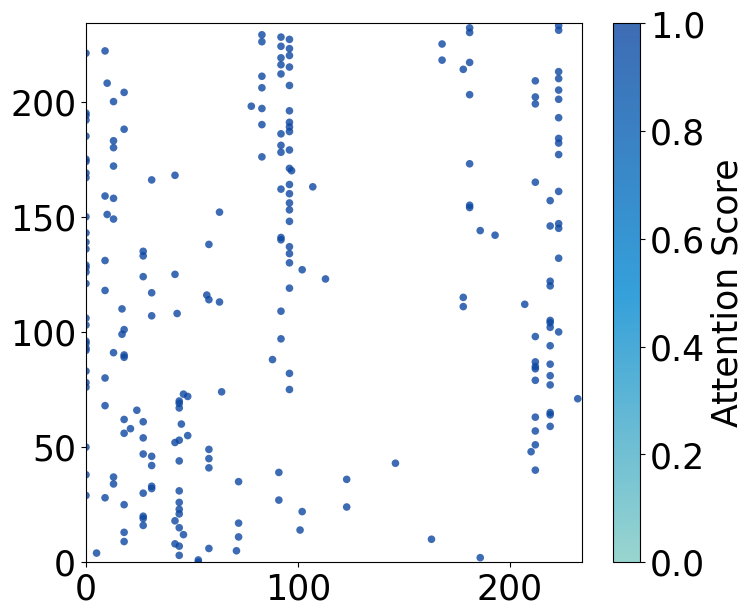}
        \caption*{(a) $HH$ Attention Score (Trained)}
    \end{minipage}
    \hfill
    \begin{minipage}[b]{0.48\textwidth}
        \centering
        \includegraphics[width=\textwidth]{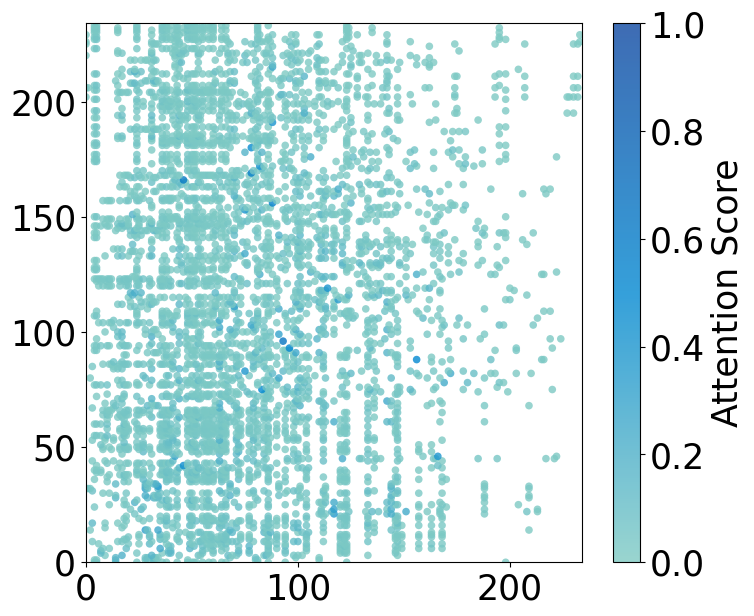}
        \caption*{(b) $HH$ Attention Score (Untrained)}
    \end{minipage}
    % \vspace{0.5cm}
    \begin{minipage}[b]{0.48\textwidth}
        \centering
        \includegraphics[width=\textwidth]{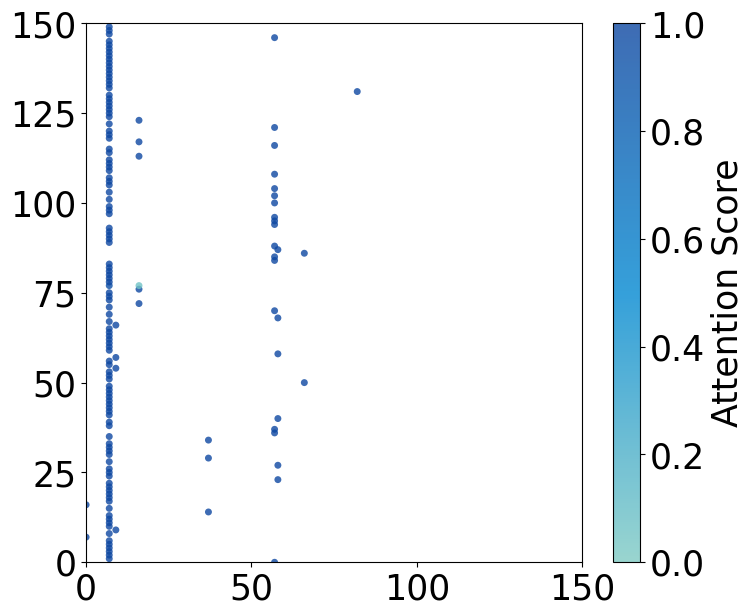}
        \caption*{(c) $2b2j$ Attention Score (Trained)}
    \end{minipage}
    \hfill
    \begin{minipage}[b]{0.48\textwidth}
        \centering
        \includegraphics[width=\textwidth]{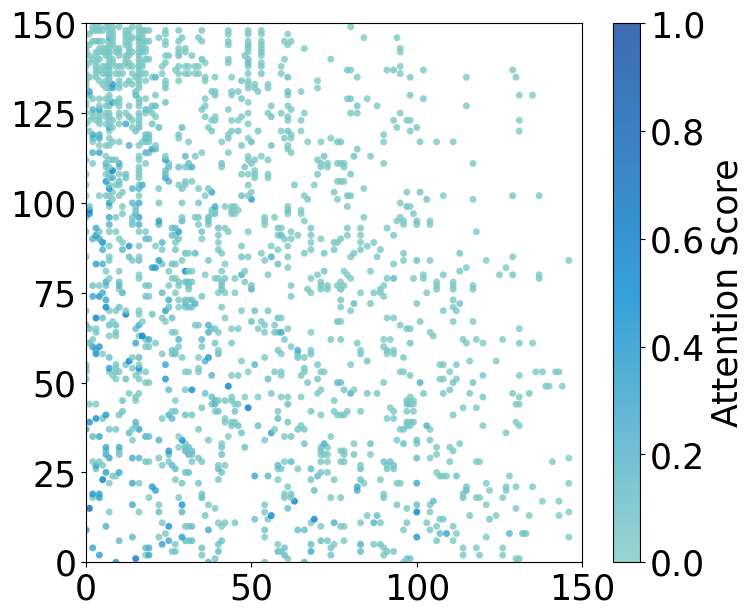}
        \caption*{(d) $2b2j$ Attention Score (Untrained)}
    \end{minipage}
    \caption{The attention score (elements in attention matrix) for $HH$ (upper panels) and $2b2j$ (lower panels) processes after (left panels) and before (right panels) training. Note that we have removed the attention score below 0.01 to make the plot clearer.}
    \label{fig:attention-heatmaps-combined}
\end{figure}

Before we conclude, in this section, we also would like to discuss the interpretability of the {\it EvenT} model to understand the internal connections of the model. As we have mentioned previously, the outstanding performance of Transformer based models is mainly attributed to the attention mechanism. Hence, we will mainly focus on the visualization of the attention mechanism of the {\it EvenT} model following the method in~\cite{Wang:2024rup} using the attention matrix in~\autoref{equ:attention_matrix}. We visualized the attention scores from the first attention head of the last particle attention block. The results are shown in~\autoref{fig:attention-heatmaps-combined} which presents the attention score (element of the attention matrix) of $HH$ (upper panels) and $2b2j$ (lower panels) processes. As a comparison, we present the attention matrix after (left panels) and before (right panels) training together. In order to make the plot clear, we ignore the attention score that is lower than 0.01. It is clear that the attention score is more concentrated for both $HH$ and $2b2j$ processes with the training which strongly indicates that the model does learn the important relationship between particles. Further, the model will pay attention to different part in the particle space for different processes, which will be the key for the model to distinguish different processes.

\begin{figure}[!tbp]
    \centering
    \begin{minipage}[b]{0.48\textwidth}
        \centering
        \includegraphics[width=\textwidth]{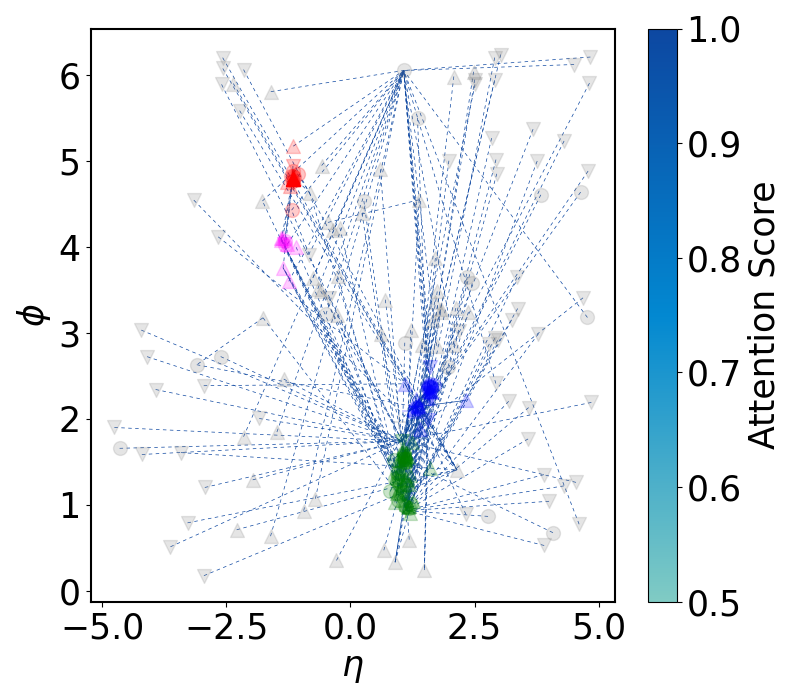}
        \caption*{(a) $HH$ Trained}
    \end{minipage}
    \begin{minipage}[b]{0.48\textwidth}
        \centering
        \includegraphics[width=\textwidth]{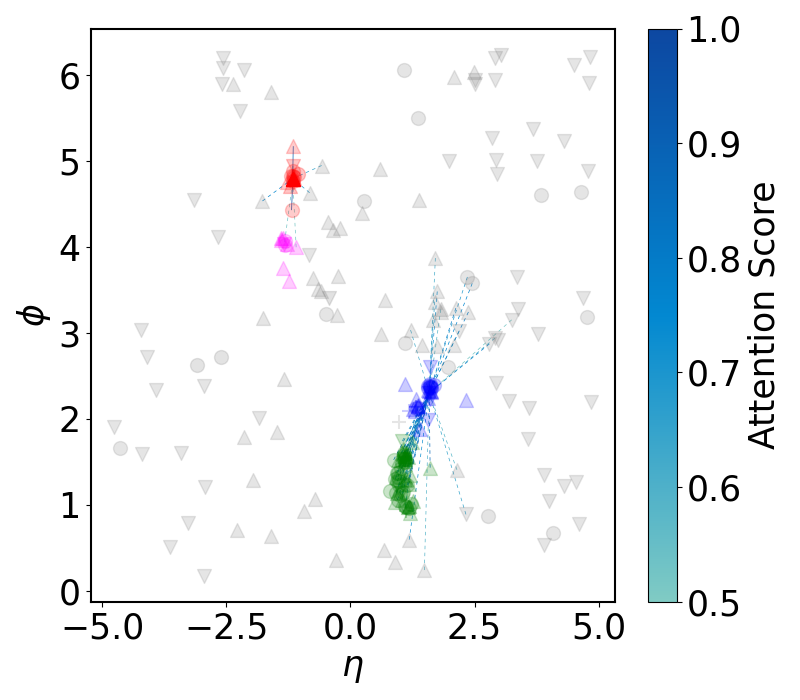}
        \caption*{(b) $HH$ Untrained}
    \end{minipage}
    % \vspace{0.5cm}
    \begin{minipage}[b]{0.48\textwidth}
        \centering
        \includegraphics[width=\textwidth]{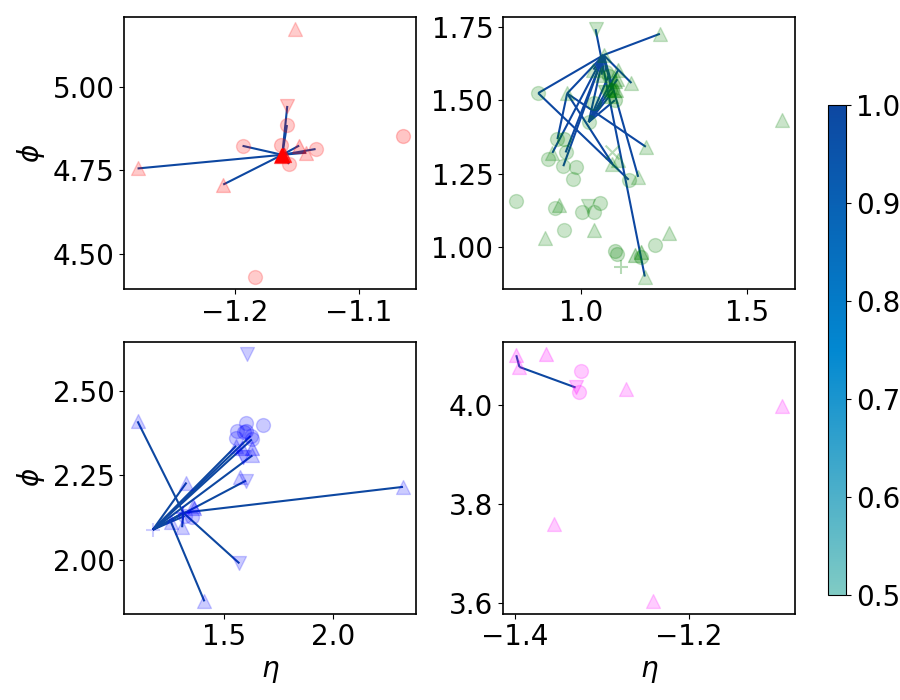}
        \caption*{(c) $HH$ Trained Zoom-in}
    \end{minipage}
    \begin{minipage}[b]{0.48\textwidth}
        \centering
        \includegraphics[width=\textwidth]{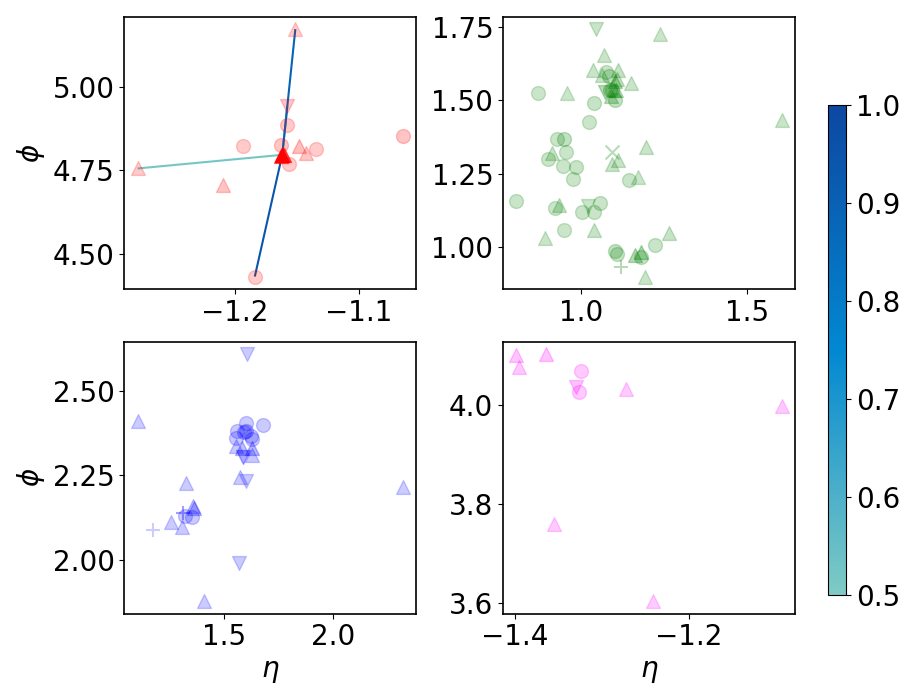}
        \caption*{(d) $HH$ Untrained Zoom-in}
    \end{minipage}
    \caption{Attention graphs for $HH$ processes before and after model training. The lower panels show the connections within a single jet.}
    \label{fig:attention-graph-comparison-HH}
\end{figure}

\begin{figure}[!tbp]
    \centering
    \begin{minipage}[b]{0.48\textwidth}
        \centering
        \includegraphics[width=\textwidth]{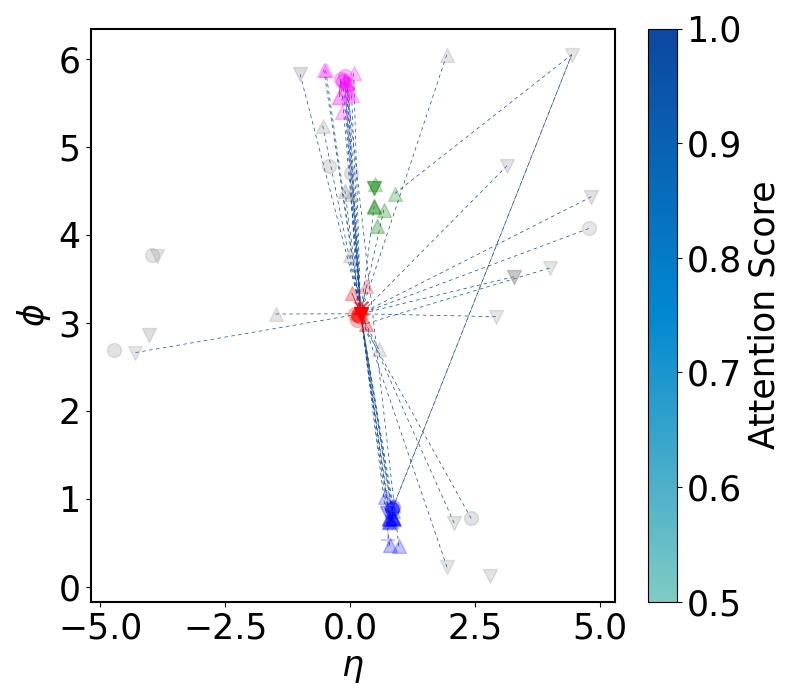}
        \caption*{(a) $2b2j$ Trained}
    \end{minipage}
    \begin{minipage}[b]{0.48\textwidth}
        \centering
        \includegraphics[width=\textwidth]{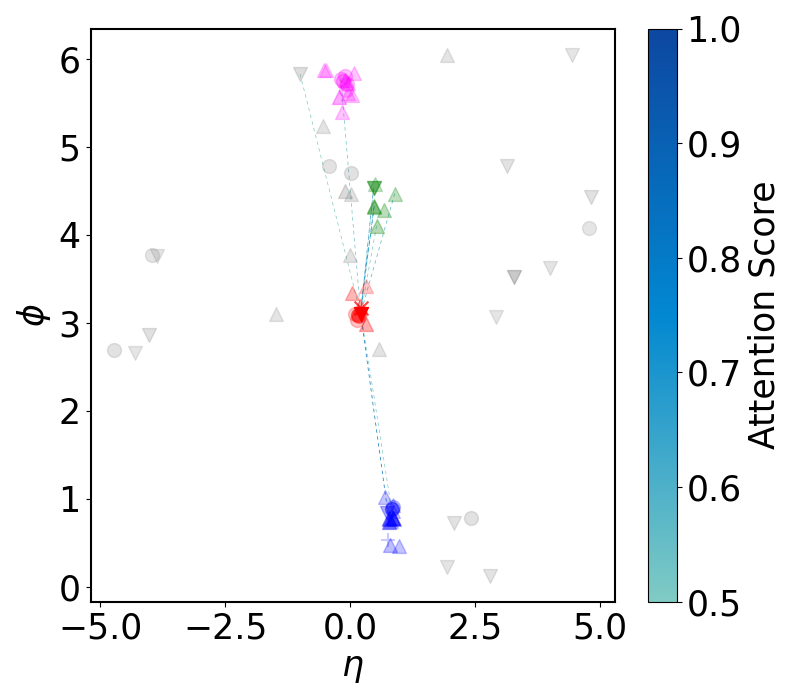}
        \caption*{(b) $2b2j$ Untrained}
    \end{minipage}
    % \vspace{0.5cm}
    \begin{minipage}[b]{0.48\textwidth}
        \centering
        \includegraphics[width=\textwidth]{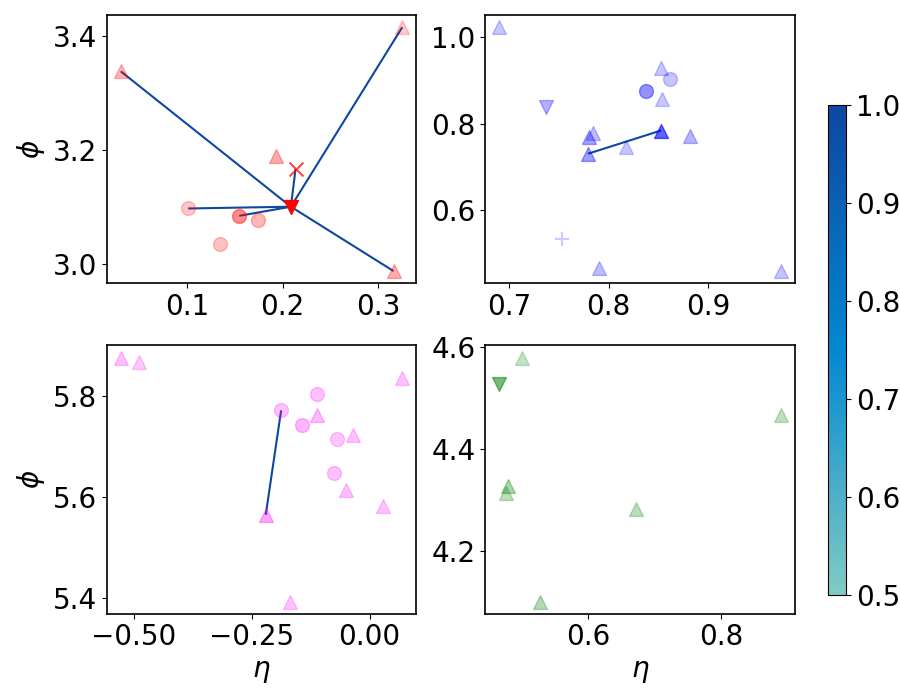}
        \caption*{(c) $2b2j$ Trained Zoom-in}
    \end{minipage}
    \begin{minipage}[b]{0.48\textwidth}
        \centering
        \includegraphics[width=\textwidth]{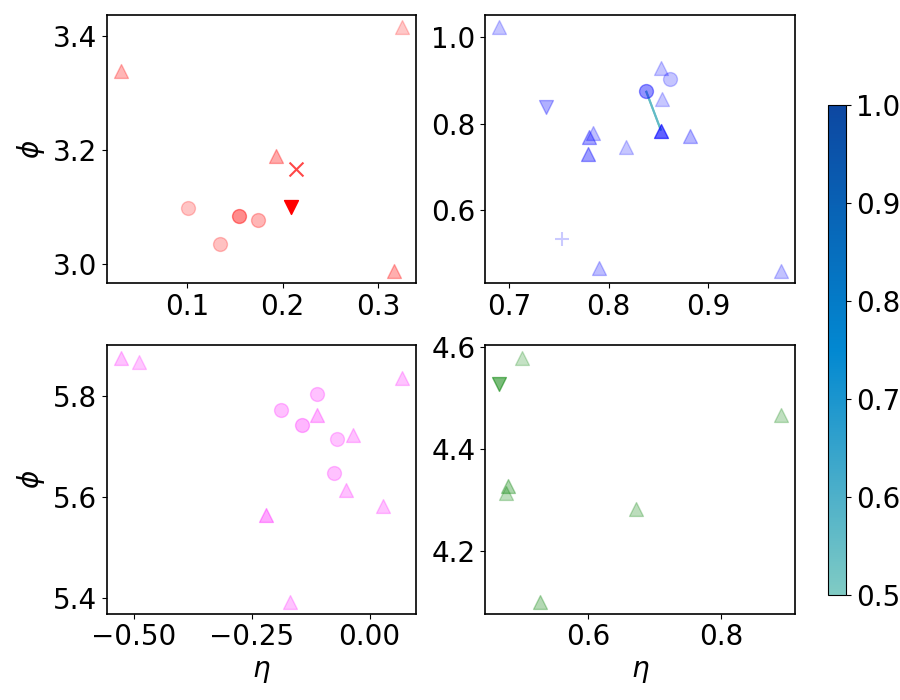}
        \caption*{(d) $2b2j$ Untrained Zoom-in}
    \end{minipage}
    \caption{Attention graphs for $2b2j$ processes before and after model training. The lower panels show the connections within a single jet.}
    \label{fig:attention-graph-comparison-2b2j}
\end{figure}

To incorporate the spatial position of particles, particle types, and the jets to which they belong during visualization, we further visualize the \textit{Particle Attention Graph}, focusing on the final layer of the P-MHA module. Each particle is represented as a point in the $\eta$-$\phi$ plane, as shown in~\autoref{fig:attention-graph-comparison-HH} and~\autoref{fig:attention-graph-comparison-2b2j} for $HH$  and $2b2j$ processes respectively. Different marker shapes are used to distinguish particle types: $\times$ for mesons, $\triangle$ for charged hadrons, $\blacktriangledown$ for neutral hadrons, $\bullet$ for photons, and $+$ for electrons. Particles within the same jet, which is reconstructed using anti-$k_t$ algorithm, are shown in the same color, while particles outside any jet are shown in gray. The opacity of each point is proportional to the transverse momentum $p_T$ of the corresponding particle. The color of the connecting lines represents the attention score between particles. Solid lines indicate connections between particles inside jets, while dashed lines indicate connections involving out-of-jet particles. Note that to make the plot clear without full of low-weight connections, we only include the connections with weight higher than $0.5$.

In the upper panels of both~\autoref{fig:attention-graph-comparison-HH} and \autoref{fig:attention-graph-comparison-2b2j}, the connections in the whole $\eta$-$\phi$ plane are shown after (upper left) and before (upper right) the training for $HH$ and $2b2j$ processes respectively. Similar to that in~\autoref{fig:attention-heatmaps-combined}, the connections among particles become stronger after the training for both $HH$ and $2b2j$ processes. To be more clear about the connections that are established during the training, for each process, we also provide the corresponding zoom-in view of the attention graph in the lower panels of~\autoref{fig:attention-graph-comparison-HH} and \autoref{fig:attention-graph-comparison-2b2j}. It is clear that the training also strengthen the connections within jets.
By comparing these visualized attention graph, we conclude that the trained model is capable of focusing on the important particle pairs in an event and learning relationships within and between jets. This capability underpins the model's outstanding performance.

\section{Conclusions}
\label{sec: conclusions}

In this work, we employ a deep neural network architecture based on {\it ParT} to enhance the sensitivity of the Higgs self-coupling measurement through $HH\to b\bar{b}b\bar{b}$ at the LHC which suffers from complex QCD backgrounds. With the help of the attention mechanism, the model can focus on important relationships among input features, enabling robust event classification when trained on full event-level information. The performance of such classifier is evaluated across 9 event categories achieving $\text{AUC}\approx 0.9$ for most categories, significantly distinguishing one process from the rest.

The studies in this work demonstrate the possibility of using the {\it ParT} model beyond the jet-tagging task. By treating the entire event as a single fat jet and modifying the input layer to accommodate the full event features, the model achieves high classification accuracy while circumventing the error-prone explicit jet-pairing process inherent in traditional event reconstruction. Such approach streamlines analysis by directly utilizing the full event information.

By applying the model to $HH\to b\bar{b}b\bar{b}$ studies at the HL-LHC, the model constrains the Higgs self-coupling $\kappa_\lambda$ to $(-0.53,\,6.01)$ at 68\% CL, which achieves around 40\% improvement in precision compared with the traditional cut-based approach. The comparison against other ML methods further highlights the advantages of the Transformer-based architecture, particularly in capturing the correlations within high-dimensional event data. Further, the attention mechanism naturally provide the interpretability of the model which can also help us understanding what are the most important features and correlations when trying to separate different processes.

\acknowledgments
The authors gratefully acknowledge Huilin Qu, Congqiao Li, Sitian Qian and Kun Wang for their insightful discussion on the {\it ParT}, as well as Huifang Lv for her valuable contributions during the early stages of this project. This work is supported by the National Natural Science Foundation of China (NNSFC) under grant No.~12305112. The authors gratefully acknowledge the valuable discussions and insights provided by the members of the China Collaboration of Precision Testing and New Physics (CPTNP).

\bibliographystyle{bibsty}
\bibliography{references}

\providecommand{\href}[2]{#2}\begingroup\raggedright\begin{thebibliography}{100}

\bibitem{ATLAS:2012yve}
{\scshape ATLAS} collaboration, \emph{{Observation of a new particle in the
  search for the Standard Model Higgs boson with the ATLAS detector at the
  LHC}}, \href{https://doi.org/10.1016/j.physletb.2012.08.020}{\emph{Phys.
  Lett. B} {\bfseries 716} (2012) 1}
  [\href{https://arxiv.org/abs/1207.7214}{{\ttfamily 1207.7214}}].

\bibitem{CMS:2012qbp}
{\scshape CMS} collaboration, \emph{{Observation of a New Boson at a Mass of
  125 GeV with the CMS Experiment at the LHC}},
  \href{https://doi.org/10.1016/j.physletb.2012.08.021}{\emph{Phys. Lett. B}
  {\bfseries 716} (2012) 30} [\href{https://arxiv.org/abs/1207.7235}{{\ttfamily
  1207.7235}}].

\bibitem{CMS:2020xrn}
{\scshape CMS} collaboration, \emph{{A measurement of the Higgs boson mass in
  the diphoton decay channel}},
  \href{https://doi.org/10.1016/j.physletb.2020.135425}{\emph{Phys. Lett. B}
  {\bfseries 805} (2020) 135425}
  [\href{https://arxiv.org/abs/2002.06398}{{\ttfamily 2002.06398}}].

\bibitem{ATLAS:2023oaq}
{\scshape ATLAS} collaboration, \emph{{Combined Measurement of the Higgs Boson
  Mass from the H\textrightarrow{}\ensuremath{\gamma}\ensuremath{\gamma} and
  H\textrightarrow{}ZZ*\textrightarrow{}4\ensuremath{\ell} Decay Channels with
  the ATLAS Detector Using s=7, 8, and 13~TeV pp Collision Data}},
  \href{https://doi.org/10.1103/PhysRevLett.131.251802}{\emph{Phys. Rev. Lett.}
  {\bfseries 131} (2023) 251802}
  [\href{https://arxiv.org/abs/2308.04775}{{\ttfamily 2308.04775}}].

\bibitem{CMS:2022ley}
{\scshape CMS} collaboration, \emph{{Measurement of the Higgs boson width and
  evidence of its off-shell contributions to ZZ production}},
  \href{https://doi.org/10.1038/s41567-022-01682-0}{\emph{Nature Phys.}
  {\bfseries 18} (2022) 1329}
  [\href{https://arxiv.org/abs/2202.06923}{{\ttfamily 2202.06923}}].

\bibitem{ATLAS:2022vkf}
{\scshape ATLAS} collaboration, \emph{{A detailed map of Higgs boson
  interactions by the ATLAS experiment ten years after the discovery}},
  \href{https://doi.org/10.1038/s41586-022-04893-w}{\emph{Nature} {\bfseries
  607} (2022) 52} [\href{https://arxiv.org/abs/2207.00092}{{\ttfamily
  2207.00092}}].

\bibitem{CMS:2022dwd}
{\scshape CMS} collaboration, \emph{{A portrait of the Higgs boson by the CMS
  experiment ten years after the discovery.}},
  \href{https://doi.org/10.1038/s41586-022-04892-x}{\emph{Nature} {\bfseries
  607} (2022) 60} [\href{https://arxiv.org/abs/2207.00043}{{\ttfamily
  2207.00043}}].

\bibitem{Abouabid:2024gms}
H.~Abouabid et~al., \emph{{HHH whitepaper}},
  \href{https://doi.org/10.1140/epjc/s10052-024-13376-3}{\emph{Eur. Phys. J. C}
  {\bfseries 84} (2024) 1183}
  [\href{https://arxiv.org/abs/2407.03015}{{\ttfamily 2407.03015}}].

\bibitem{Lu:2015jza}
C.-T.~Lu, J.~Chang, K.~Cheung and J.S.~Lee, \emph{{An exploratory study of
  Higgs-boson pair production}},
  \href{https://doi.org/10.1007/JHEP08(2015)133}{\emph{JHEP} {\bfseries 08}
  (2015) 133} [\href{https://arxiv.org/abs/1505.00957}{{\ttfamily
  1505.00957}}].

\bibitem{Kling:2016lay}
F.~Kling, T.~Plehn and P.~Schichtel, \emph{{Maximizing the significance in
  Higgs boson pair analyses}},
  \href{https://doi.org/10.1103/PhysRevD.95.035026}{\emph{Phys. Rev. D}
  {\bfseries 95} (2017) 035026}
  [\href{https://arxiv.org/abs/1607.07441}{{\ttfamily 1607.07441}}].

\bibitem{DiMicco:2019ngk}
J.~Alison et~al., \emph{{Higgs boson potential at colliders: Status and
  perspectives}}, \href{https://doi.org/10.1016/j.revip.2020.100045}{\emph{Rev.
  Phys.} {\bfseries 5} (2020) 100045}
  [\href{https://arxiv.org/abs/1910.00012}{{\ttfamily 1910.00012}}].

\bibitem{Baur:2003gp}
U.~Baur, T.~Plehn and D.L.~Rainwater, \emph{{Probing the Higgs selfcoupling at
  hadron colliders using rare decays}},
  \href{https://doi.org/10.1103/PhysRevD.69.053004}{\emph{Phys. Rev. D}
  {\bfseries 69} (2004) 053004}
  [\href{https://arxiv.org/abs/hep-ph/0310056}{{\ttfamily hep-ph/0310056}}].

\bibitem{Degrassi:2021uik}
G.~Degrassi, B.~Di~Micco, P.P.~Giardino and E.~Rossi, \emph{{Higgs boson
  self-coupling constraints from single Higgs, double Higgs and Electroweak
  measurements}},
  \href{https://doi.org/10.1016/j.physletb.2021.136307}{\emph{Phys. Lett. B}
  {\bfseries 817} (2021) 136307}
  [\href{https://arxiv.org/abs/2102.07651}{{\ttfamily 2102.07651}}].

\bibitem{Bahl:2023lck}
H.~Bahl, J.~Braathen, M.~Gabelmann and G.R.~Weiglein, \emph{{Precise
  predictions for the trilinear Higgs self-coupling in the Standard Model and
  beyond}}, \href{https://doi.org/10.22323/1.449.0407}{\emph{PoS} {\bfseries
  EPS-HEP2023} (2024) 407} [\href{https://arxiv.org/abs/2311.01134}{{\ttfamily
  2311.01134}}].

\bibitem{Arco:2022lai}
F.~Arco, S.~Heinemeyer, M.~M\"uhlleitner and K.~Radchenko, \emph{{Sensitivity
  to triple Higgs couplings via di-Higgs production in the 2HDM at the
  (HL-)LHC}}, \href{https://doi.org/10.1140/epjc/s10052-023-12193-4}{\emph{Eur.
  Phys. J. C} {\bfseries 83} (2023) 1019}
  [\href{https://arxiv.org/abs/2212.11242}{{\ttfamily 2212.11242}}].

\bibitem{Papaefstathiou:2012qe}
A.~Papaefstathiou, L.L.~Yang and J.~Zurita, \emph{{Higgs boson pair production
  at the LHC in the $b \bar{b} W^+ W^-$ channel}},
  \href{https://doi.org/10.1103/PhysRevD.87.011301}{\emph{Phys. Rev. D}
  {\bfseries 87} (2013) 011301}
  [\href{https://arxiv.org/abs/1209.1489}{{\ttfamily 1209.1489}}].

\bibitem{Baur:2002rb}
U.~Baur, T.~Plehn and D.L.~Rainwater, \emph{{Measuring the Higgs Boson Self
  Coupling at the LHC and Finite Top Mass Matrix Elements}},
  \href{https://doi.org/10.1103/PhysRevLett.89.151801}{\emph{Phys. Rev. Lett.}
  {\bfseries 89} (2002) 151801}
  [\href{https://arxiv.org/abs/hep-ph/0206024}{{\ttfamily hep-ph/0206024}}].

\bibitem{Baglio:2012np}
J.~Baglio, A.~Djouadi, R.~Gr\"ober, M.M.~M\"uhlleitner, J.~Quevillon and
  M.~Spira, \emph{{The measurement of the Higgs self-coupling at the LHC:
  theoretical status}},
  \href{https://doi.org/10.1007/JHEP04(2013)151}{\emph{JHEP} {\bfseries 04}
  (2013) 151} [\href{https://arxiv.org/abs/1212.5581}{{\ttfamily 1212.5581}}].

\bibitem{Dolan:2012ac}
M.J.~Dolan, C.~Englert and M.~Spannowsky, \emph{{New Physics in LHC Higgs boson
  pair production}},
  \href{https://doi.org/10.1103/PhysRevD.87.055002}{\emph{Phys. Rev. D}
  {\bfseries 87} (2013) 055002}
  [\href{https://arxiv.org/abs/1210.8166}{{\ttfamily 1210.8166}}].

\bibitem{Durieux:2022hbu}
G.~Durieux, M.~McCullough and E.~Salvioni, \emph{{Charting the Higgs
  self-coupling boundaries}},
  \href{https://doi.org/10.1007/JHEP12(2022)148}{\emph{JHEP} {\bfseries 12}
  (2022) 148} [\href{https://arxiv.org/abs/2209.00666}{{\ttfamily
  2209.00666}}].

\bibitem{Cheung:2020xij}
K.~Cheung, A.~Jueid, C.-T.~Lu, J.~Song and Y.W.~Yoon, \emph{{Disentangling new
  physics effects on nonresonant Higgs boson pair production from gluon
  fusion}}, \href{https://doi.org/10.1103/PhysRevD.103.015019}{\emph{Phys. Rev.
  D} {\bfseries 103} (2021) 015019}
  [\href{https://arxiv.org/abs/2003.11043}{{\ttfamily 2003.11043}}].

\bibitem{Chang:2019vez}
S.~Chang and M.A.~Luty, \emph{{The Higgs Trilinear Coupling and the Scale of
  New Physics}}, \href{https://doi.org/10.1007/JHEP03(2020)140}{\emph{JHEP}
  {\bfseries 03} (2020) 140}
  [\href{https://arxiv.org/abs/1902.05556}{{\ttfamily 1902.05556}}].

\bibitem{Gupta:2013zza}
R.S.~Gupta, H.~Rzehak and J.D.~Wells, \emph{{How well do we need to measure the
  Higgs boson mass and self-coupling?}},
  \href{https://doi.org/10.1103/PhysRevD.88.055024}{\emph{Phys. Rev. D}
  {\bfseries 88} (2013) 055024}
  [\href{https://arxiv.org/abs/1305.6397}{{\ttfamily 1305.6397}}].

\bibitem{Bhattiprolu:2024tsq}
P.N.~Bhattiprolu and J.D.~Wells, \emph{{A sensitivity target for an impactful
  Higgs boson self coupling measurement}},
  \href{https://doi.org/10.1016/j.physletb.2025.139263}{\emph{Phys. Lett. B}
  {\bfseries 861} (2025) 139263}
  [\href{https://arxiv.org/abs/2407.11847}{{\ttfamily 2407.11847}}].

\bibitem{Dawson:2015oha}
S.~Dawson, A.~Ismail and I.~Low, \emph{{What\textquoteright{}s in the loop? The
  anatomy of double Higgs production}},
  \href{https://doi.org/10.1103/PhysRevD.91.115008}{\emph{Phys. Rev. D}
  {\bfseries 91} (2015) 115008}
  [\href{https://arxiv.org/abs/1504.05596}{{\ttfamily 1504.05596}}].

\bibitem{Banerjee:2016nzb}
S.~Banerjee, B.~Batell and M.~Spannowsky, \emph{{Invisible decays in Higgs
  boson pair production}},
  \href{https://doi.org/10.1103/PhysRevD.95.035009}{\emph{Phys. Rev. D}
  {\bfseries 95} (2017) 035009}
  [\href{https://arxiv.org/abs/1608.08601}{{\ttfamily 1608.08601}}].

\bibitem{Alasfar:2023xpc}
L.~Alasfar et~al., \emph{{Effective Field Theory descriptions of Higgs boson
  pair production}},
  \href{https://doi.org/10.21468/SciPostPhysCommRep.2}{\emph{SciPost Phys.
  Comm. Rep.} {\bfseries 2024} (2024) 2}
  [\href{https://arxiv.org/abs/2304.01968}{{\ttfamily 2304.01968}}].

\bibitem{Goertz:2014qta}
F.~Goertz, A.~Papaefstathiou, L.L.~Yang and J.~Zurita, \emph{{Higgs boson pair
  production in the D=6 extension of the SM}},
  \href{https://doi.org/10.1007/JHEP04(2015)167}{\emph{JHEP} {\bfseries 04}
  (2015) 167} [\href{https://arxiv.org/abs/1410.3471}{{\ttfamily 1410.3471}}].

\bibitem{Azatov:2015oxa}
A.~Azatov, R.~Contino, G.~Panico and M.~Son, \emph{{Effective field theory
  analysis of double Higgs boson production via gluon fusion}},
  \href{https://doi.org/10.1103/PhysRevD.92.035001}{\emph{Phys. Rev. D}
  {\bfseries 92} (2015) 035001}
  [\href{https://arxiv.org/abs/1502.00539}{{\ttfamily 1502.00539}}].

\bibitem{Cao:2015oaa}
Q.-H.~Cao, B.~Yan, D.-M.~Zhang and H.~Zhang, \emph{{Resolving the Degeneracy in
  Single Higgs Production with Higgs Pair Production}},
  \href{https://doi.org/10.1016/j.physletb.2015.11.045}{\emph{Phys. Lett. B}
  {\bfseries 752} (2016) 285}
  [\href{https://arxiv.org/abs/1508.06512}{{\ttfamily 1508.06512}}].

\bibitem{Li:2019uyy}
G.~Li, L.-X.~Xu, B.~Yan and C.P.~Yuan, \emph{{Resolving the degeneracy in top
  quark Yukawa coupling with Higgs pair production}},
  \href{https://doi.org/10.1016/j.physletb.2019.135070}{\emph{Phys. Lett. B}
  {\bfseries 800} (2020) 135070}
  [\href{https://arxiv.org/abs/1904.12006}{{\ttfamily 1904.12006}}].

\bibitem{Liu:2018peg}
T.~Liu, K.-F.~Lyu, J.~Ren and H.X.~Zhu, \emph{{Probing the quartic Higgs boson
  self-interaction}},
  \href{https://doi.org/10.1103/PhysRevD.98.093004}{\emph{Phys. Rev. D}
  {\bfseries 98} (2018) 093004}
  [\href{https://arxiv.org/abs/1803.04359}{{\ttfamily 1803.04359}}].

\bibitem{Bizon:2018syu}
W.~Bizo\'n, U.~Haisch and L.~Rottoli, \emph{{Constraints on the quartic Higgs
  self-coupling from double-Higgs production at future hadron colliders}},
  \href{https://doi.org/10.1007/JHEP02(2024)170}{\emph{JHEP} {\bfseries 10}
  (2019) 267} [\href{https://arxiv.org/abs/1810.04665}{{\ttfamily
  1810.04665}}].

\bibitem{Chen:2015gva}
C.-Y.~Chen, Q.-S.~Yan, X.~Zhao, Y.-M.~Zhong and Z.~Zhao, \emph{{Probing
  triple-Higgs productions via 4b2\ensuremath{\gamma} decay channel at a 100
  TeV hadron collider}},
  \href{https://doi.org/10.1103/PhysRevD.93.013007}{\emph{Phys. Rev. D}
  {\bfseries 93} (2016) 013007}
  [\href{https://arxiv.org/abs/1510.04013}{{\ttfamily 1510.04013}}].

\bibitem{Fuks:2017zkg}
B.~Fuks, J.H.~Kim and S.J.~Lee, \emph{{Scrutinizing the Higgs quartic coupling
  at a future 100 TeV proton\textendash{}proton collider with taus and
  b-jets}}, \href{https://doi.org/10.1016/j.physletb.2017.05.075}{\emph{Phys.
  Lett. B} {\bfseries 771} (2017) 354}
  [\href{https://arxiv.org/abs/1704.04298}{{\ttfamily 1704.04298}}].

\bibitem{Chiesa:2020awd}
M.~Chiesa, F.~Maltoni, L.~Mantani, B.~Mele, F.~Piccinini and X.~Zhao,
  \emph{{Measuring the quartic Higgs self-coupling at a multi-TeV muon
  collider}}, \href{https://doi.org/10.1007/JHEP09(2020)098}{\emph{JHEP}
  {\bfseries 09} (2020) 098}
  [\href{https://arxiv.org/abs/2003.13628}{{\ttfamily 2003.13628}}].

\bibitem{Torndal:2023mmr}
J.M.~Torndal, J.~List, D.~Ntounis and C.~Vernieri, \emph{{Higgs self-coupling
  measurement at future~$e^+e^-$ colliders}},
  \href{https://doi.org/10.22323/1.449.0406}{\emph{PoS} {\bfseries EPS-HEP2023}
  (2024) 406} [\href{https://arxiv.org/abs/2311.16774}{{\ttfamily
  2311.16774}}].

\bibitem{Torndal:2023fky}
J.M.~Torndal and J.~List, \emph{{Higgs self-coupling measurement at the
  International Linear Collider}},  in \emph{{International Workshop on Future
  Linear Colliders}}, 7, 2023
  [\href{https://arxiv.org/abs/2307.16515}{{\ttfamily 2307.16515}}].

\bibitem{Davies:2024kvt}
J.~Davies, \emph{{Higgs pair production at NNLO}},
  \href{https://doi.org/10.22323/1.467.0015}{\emph{PoS} {\bfseries LL2024}
  (2024) 015} [\href{https://arxiv.org/abs/2407.08264}{{\ttfamily
  2407.08264}}].

\bibitem{Zhang:2024rix}
H.~Zhang, K.~Sch\"onwald, M.~Steinhauser and J.~Davies, \emph{{Electroweak
  corrections to gg -\ensuremath{>} HH: Factorizable contributions}},
  \href{https://doi.org/10.22323/1.467.0014}{\emph{PoS} {\bfseries LL2024}
  (2024) 014} [\href{https://arxiv.org/abs/2407.05787}{{\ttfamily
  2407.05787}}].

\bibitem{Heinrich:2024dnz}
G.~Heinrich, S.~Jones, M.~Kerner, T.~Stone and A.~Vestner, \emph{{Electroweak
  corrections to Higgs boson pair production: the top-Yukawa and self-coupling
  contributions}}, \href{https://doi.org/10.1007/JHEP11(2024)040}{\emph{JHEP}
  {\bfseries 11} (2024) 040}
  [\href{https://arxiv.org/abs/2407.04653}{{\ttfamily 2407.04653}}].

\bibitem{Carvalho:2015ttv}
A.~Carvalho, M.~Dall'Osso, T.~Dorigo, F.~Goertz, C.A.~Gottardo and M.~Tosi,
  \emph{{Higgs Pair Production: Choosing Benchmarks With Cluster Analysis}},
  \href{https://doi.org/10.1007/JHEP04(2016)126}{\emph{JHEP} {\bfseries 04}
  (2016) 126} [\href{https://arxiv.org/abs/1507.02245}{{\ttfamily
  1507.02245}}].

\bibitem{Kim:2018cxf}
J.H.~Kim, K.~Kong, K.T.~Matchev and M.~Park, \emph{{Probing the Triple Higgs
  Self-Interaction at the Large Hadron Collider}},
  \href{https://doi.org/10.1103/PhysRevLett.122.091801}{\emph{Phys. Rev. Lett.}
  {\bfseries 122} (2019) 091801}
  [\href{https://arxiv.org/abs/1807.11498}{{\ttfamily 1807.11498}}].

\bibitem{Bizon:2024juq}
W.~Bizo\'n, U.~Haisch, L.~Rottoli, Z.~Gillis, B.~Moser and P.~Windischhofer,
  \emph{{Addendum to: Constraints on the quartic Higgs self-coupling from
  double-Higgs production at future hadron colliders [JHEP~10~(2019)~267]}},
  \href{https://doi.org/10.1007/JHEP02(2024)170}{\emph{JHEP} {\bfseries 02}
  (2024) 170} [\href{https://arxiv.org/abs/2402.03463}{{\ttfamily
  2402.03463}}].

\bibitem{Nakamura:2017irk}
K.~Nakamura, K.~Nishiwaki, K.-y.~Oda, S.C.~Park and Y.~Yamamoto,
  \emph{{Di-higgs enhancement by neutral scalar as probe of new colored
  sector}}, \href{https://doi.org/10.1140/epjc/s10052-017-4835-4}{\emph{Eur.
  Phys. J. C} {\bfseries 77} (2017) 273}
  [\href{https://arxiv.org/abs/1701.06137}{{\ttfamily 1701.06137}}].

\bibitem{Chen:2019lzz}
L.-B.~Chen, H.T.~Li, H.-S.~Shao and J.~Wang, \emph{{Higgs boson pair production
  via gluon fusion at N$^3$LO in QCD}},
  \href{https://doi.org/10.1016/j.physletb.2020.135292}{\emph{Phys. Lett. B}
  {\bfseries 803} (2020) 135292}
  [\href{https://arxiv.org/abs/1909.06808}{{\ttfamily 1909.06808}}].

\bibitem{Baglio:2020ini}
J.~Baglio, F.~Campanario, S.~Glaus, M.~M\"uhlleitner, J.~Ronca, M.~Spira
  et~al., \emph{{Higgs-Pair Production via Gluon Fusion at Hadron Colliders:
  NLO QCD Corrections}},
  \href{https://doi.org/10.1007/JHEP04(2020)181}{\emph{JHEP} {\bfseries 04}
  (2020) 181} [\href{https://arxiv.org/abs/2003.03227}{{\ttfamily
  2003.03227}}].

\bibitem{Baglio:2018lrj}
J.~Baglio, F.~Campanario, S.~Glaus, M.~M\"uhlleitner, M.~Spira and
  J.~Streicher, \emph{{Gluon fusion into Higgs pairs at NLO QCD and the top
  mass scheme}},
  \href{https://doi.org/10.1140/epjc/s10052-019-6973-3}{\emph{Eur. Phys. J. C}
  {\bfseries 79} (2019) 459}
  [\href{https://arxiv.org/abs/1811.05692}{{\ttfamily 1811.05692}}].

\bibitem{Borowka:2016ehy}
S.~Borowka, N.~Greiner, G.~Heinrich, S.P.~Jones, M.~Kerner, J.~Schlenk et~al.,
  \emph{{Higgs Boson Pair Production in Gluon Fusion at Next-to-Leading Order
  with Full Top-Quark Mass Dependence}},
  \href{https://doi.org/10.1103/PhysRevLett.117.079901}{\emph{Phys. Rev. Lett.}
  {\bfseries 117} (2016) 012001}
  [\href{https://arxiv.org/abs/1604.06447}{{\ttfamily 1604.06447}}].

\bibitem{Czurylo:2023nxf}
M.M.~Czury\l{}o, \emph{{Search for the non-resonant Vector Boson Fusion
  production of the Higgs boson pairs decaying to $b\bar{b}b\bar{b}$ final
  state using the ATLAS detector}}, Ph.D. thesis, U. Heidelberg (main), 2023.
\newblock 10.11588/heidok.00032906.

\bibitem{Cao:2015oxx}
Q.-H.~Cao, Y.~Liu and B.~Yan, \emph{{Measuring trilinear Higgs coupling in WHH
  and ZHH productions at the high-luminosity LHC}},
  \href{https://doi.org/10.1103/PhysRevD.95.073006}{\emph{Phys. Rev. D}
  {\bfseries 95} (2017) 073006}
  [\href{https://arxiv.org/abs/1511.03311}{{\ttfamily 1511.03311}}].

\bibitem{Chai:2022zeq}
K.~Chai, J.-H.~Yu and H.~Zhang, \emph{{Investigating Higgs self-interaction
  through di-Higgs plus jet production at a 100~TeV hadron collider}},
  \href{https://doi.org/10.1103/PhysRevD.107.055031}{\emph{Phys. Rev. D}
  {\bfseries 107} (2023) 055031}
  [\href{https://arxiv.org/abs/2210.14929}{{\ttfamily 2210.14929}}].

\bibitem{Dolan:2013rja}
M.J.~Dolan, C.~Englert, N.~Greiner and M.~Spannowsky, \emph{{Further on up the
  road: $hhjj$ production at the LHC}},
  \href{https://doi.org/10.1103/PhysRevLett.112.101802}{\emph{Phys. Rev. Lett.}
  {\bfseries 112} (2014) 101802}
  [\href{https://arxiv.org/abs/1310.1084}{{\ttfamily 1310.1084}}].

\bibitem{Abouabid:2021yvw}
H.~Abouabid, A.~Arhrib, D.~Azevedo, J.E.~Falaki, P.M.~Ferreira,
  M.~M\"uhlleitner et~al., \emph{{Benchmarking di-Higgs production in various
  extended Higgs sector models}},
  \href{https://doi.org/10.1007/JHEP09(2022)011}{\emph{JHEP} {\bfseries 09}
  (2022) 011} [\href{https://arxiv.org/abs/2112.12515}{{\ttfamily
  2112.12515}}].

\bibitem{Godunov:2014waa}
S.I.~Godunov, M.I.~Vysotsky and E.V.~Zhemchugov, \emph{{Double Higgs production
  at LHC, see-saw type II and Georgi-Machacek model}},
  \href{https://doi.org/10.1134/S1063776115030073}{\emph{J. Exp. Theor. Phys.}
  {\bfseries 120} (2015) 369}
  [\href{https://arxiv.org/abs/1408.0184}{{\ttfamily 1408.0184}}].

\bibitem{Kotwal:2015rba}
A.V.~Kotwal, S.~Chekanov and M.~Low, \emph{{Double Higgs Boson Production in
  the 4$\tau$ Channel from Resonances in Longitudinal Vector Boson Scattering
  at a 100 TeV Collider}},
  \href{https://doi.org/10.1103/PhysRevD.91.114018}{\emph{Phys. Rev. D}
  {\bfseries 91} (2015) 114018}
  [\href{https://arxiv.org/abs/1504.08042}{{\ttfamily 1504.08042}}].

\bibitem{DiLuzio:2017tfn}
L.~Di~Luzio, R.~Gr\"ober and M.~Spannowsky, \emph{{Maxi-sizing the trilinear
  Higgs self-coupling: how large could it be?}},
  \href{https://doi.org/10.1140/epjc/s10052-017-5361-0}{\emph{Eur. Phys. J. C}
  {\bfseries 77} (2017) 788}
  [\href{https://arxiv.org/abs/1704.02311}{{\ttfamily 1704.02311}}].

\bibitem{Grober:2017gut}
R.~Grober, M.~Muhlleitner and M.~Spira, \emph{{Higgs Pair Production at NLO QCD
  for CP-violating Higgs Sectors}},
  \href{https://doi.org/10.1016/j.nuclphysb.2017.10.002}{\emph{Nucl. Phys. B}
  {\bfseries 925} (2017) 1} [\href{https://arxiv.org/abs/1705.05314}{{\ttfamily
  1705.05314}}].

\bibitem{Huang:2017jws}
T.~Huang, J.M.~No, L.~Perni\'e, M.~Ramsey-Musolf, A.~Safonov, M.~Spannowsky
  et~al., \emph{{Resonant di-Higgs boson production in the $b{\bar b}WW$
  channel: Probing the electroweak phase transition at the LHC}},
  \href{https://doi.org/10.1103/PhysRevD.96.035007}{\emph{Phys. Rev. D}
  {\bfseries 96} (2017) 035007}
  [\href{https://arxiv.org/abs/1701.04442}{{\ttfamily 1701.04442}}].

\bibitem{No:2013wsa}
J.M.~No and M.~Ramsey-Musolf, \emph{{Probing the Higgs Portal at the LHC
  Through Resonant di-Higgs Production}},
  \href{https://doi.org/10.1103/PhysRevD.89.095031}{\emph{Phys. Rev. D}
  {\bfseries 89} (2014) 095031}
  [\href{https://arxiv.org/abs/1310.6035}{{\ttfamily 1310.6035}}].

\bibitem{Barger:2014taa}
V.~Barger, L.L.~Everett, C.B.~Jackson, A.D.~Peterson and G.~Shaughnessy,
  \emph{{New physics in resonant production of Higgs boson pairs}},
  \href{https://doi.org/10.1103/PhysRevLett.114.011801}{\emph{Phys. Rev. Lett.}
  {\bfseries 114} (2015) 011801}
  [\href{https://arxiv.org/abs/1408.0003}{{\ttfamily 1408.0003}}].

\bibitem{Chen:2014ask}
C.-Y.~Chen, S.~Dawson and I.M.~Lewis, \emph{{Exploring resonant di-Higgs boson
  production in the Higgs singlet model}},
  \href{https://doi.org/10.1103/PhysRevD.91.035015}{\emph{Phys. Rev. D}
  {\bfseries 91} (2015) 035015}
  [\href{https://arxiv.org/abs/1410.5488}{{\ttfamily 1410.5488}}].

\bibitem{Barducci:2019xkq}
D.~Barducci, K.~Mimasu, J.M.~No, C.~Vernieri and J.~Zurita, \emph{{Enlarging
  the scope of resonant di-Higgs searches: Hunting for Higgs-to-Higgs cascades
  in $4b$ final states at the LHC and future colliders}},
  \href{https://doi.org/10.1007/JHEP02(2020)002}{\emph{JHEP} {\bfseries 02}
  (2020) 002} [\href{https://arxiv.org/abs/1910.08574}{{\ttfamily
  1910.08574}}].

\bibitem{Heinemeyer:2024hxa}
S.~Heinemeyer, M.~M\"uhlleitner, K.~Radchenko and G.~Weiglein, \emph{{Higgs
  pair production in the 2HDM: impact of loop corrections to the trilinear
  Higgs couplings and interference effects on experimental limits}},
  \href{https://doi.org/10.1140/epjc/s10052-025-14124-x}{\emph{Eur. Phys. J. C}
  {\bfseries 85} (2025) 437}
  [\href{https://arxiv.org/abs/2403.14776}{{\ttfamily 2403.14776}}].

\bibitem{Arhrib:2024hed}
A.~Arhrib, S.~Moretti, S.~Semlali, C.H.~Shepherd-Themistocleous, Y.~Wang and
  Q.S.~Yan, \emph{{Probing a 2HDM Type-I light Higgs state via $H_{\rm SM} \to
  hh \to b\bar b\gamma \gamma$ at the LHC}},
  \href{https://arxiv.org/abs/2412.06052}{{\ttfamily 2412.06052}}.

\bibitem{Baglio:2014nea}
J.~Baglio, O.~Eberhardt, U.~Nierste and M.~Wiebusch, \emph{{Benchmarks for
  Higgs Pair Production and Heavy Higgs boson Searches in the Two-Higgs-Doublet
  Model of Type II}},
  \href{https://doi.org/10.1103/PhysRevD.90.015008}{\emph{Phys. Rev. D}
  {\bfseries 90} (2014) 015008}
  [\href{https://arxiv.org/abs/1403.1264}{{\ttfamily 1403.1264}}].

\bibitem{Hespel:2014sla}
B.~Hespel, D.~Lopez-Val and E.~Vryonidou, \emph{{Higgs pair production via
  gluon fusion in the Two-Higgs-Doublet Model}},
  \href{https://doi.org/10.1007/JHEP09(2014)124}{\emph{JHEP} {\bfseries 09}
  (2014) 124} [\href{https://arxiv.org/abs/1407.0281}{{\ttfamily 1407.0281}}].

\bibitem{Bauer:2017cov}
M.~Bauer, M.~Carena and A.~Carmona, \emph{{Higgs Pair Production as a Signal of
  Enhanced Yukawa Couplings}},
  \href{https://doi.org/10.1103/PhysRevLett.121.021801}{\emph{Phys. Rev. Lett.}
  {\bfseries 121} (2018) 021801}
  [\href{https://arxiv.org/abs/1801.00363}{{\ttfamily 1801.00363}}].

\bibitem{Arganda:2017wjh}
E.~Arganda, J.L.~D\'\i{}az-Cruz, N.~Mileo, R.A.~Morales and A.~Szynkman,
  \emph{{Search strategies for pair production of heavy Higgs bosons decaying
  invisibly at the LHC}},
  \href{https://doi.org/10.1016/j.nuclphysb.2018.02.004}{\emph{Nucl. Phys. B}
  {\bfseries 929} (2018) 171}
  [\href{https://arxiv.org/abs/1710.07254}{{\ttfamily 1710.07254}}].

\bibitem{Chen:2021pqi}
J.~Chen, T.~Li, C.-T.~Lu, Y.~Wu and C.-Y.~Yao, \emph{{Measurement of Higgs
  boson self-couplings through 2\textrightarrow{}3 vector bosons scattering in
  future muon colliders}},
  \href{https://doi.org/10.1103/PhysRevD.105.053009}{\emph{Phys. Rev. D}
  {\bfseries 105} (2022) 053009}
  [\href{https://arxiv.org/abs/2112.12507}{{\ttfamily 2112.12507}}].

\bibitem{Henning:2018kys}
B.~Henning, D.~Lombardo, M.~Riembau and F.~Riva, \emph{{Measuring Higgs
  Couplings without Higgs Bosons}},
  \href{https://doi.org/10.1103/PhysRevLett.123.181801}{\emph{Phys. Rev. Lett.}
  {\bfseries 123} (2019) 181801}
  [\href{https://arxiv.org/abs/1812.09299}{{\ttfamily 1812.09299}}].

\bibitem{McCullough:2013rea}
M.~McCullough, \emph{{An Indirect Model-Dependent Probe of the Higgs
  Self-Coupling}},
  \href{https://doi.org/10.1103/PhysRevD.90.015001}{\emph{Phys. Rev. D}
  {\bfseries 90} (2014) 015001}
  [\href{https://arxiv.org/abs/1312.3322}{{\ttfamily 1312.3322}}].

\bibitem{Gorbahn:2016uoy}
M.~Gorbahn and U.~Haisch, \emph{{Indirect probes of the trilinear Higgs
  coupling: $gg \to h$ and $h \to \gamma \gamma$}},
  \href{https://doi.org/10.1007/JHEP10(2016)094}{\emph{JHEP} {\bfseries 10}
  (2016) 094} [\href{https://arxiv.org/abs/1607.03773}{{\ttfamily
  1607.03773}}].

\bibitem{Degrassi:2016wml}
G.~Degrassi, P.P.~Giardino, F.~Maltoni and D.~Pagani, \emph{{Probing the Higgs
  self coupling via single Higgs production at the LHC}},
  \href{https://doi.org/10.1007/JHEP12(2016)080}{\emph{JHEP} {\bfseries 12}
  (2016) 080} [\href{https://arxiv.org/abs/1607.04251}{{\ttfamily
  1607.04251}}].

\bibitem{Gao:2023bll}
J.~Gao, X.-M.~Shen, G.~Wang, L.L.~Yang and B.~Zhou, \emph{{Probing the Higgs
  boson trilinear self-coupling through Higgs boson+jet production}},
  \href{https://doi.org/10.1103/PhysRevD.107.115017}{\emph{Phys. Rev. D}
  {\bfseries 107} (2023) 115017}
  [\href{https://arxiv.org/abs/2302.04160}{{\ttfamily 2302.04160}}].

\bibitem{Degrassi:2017ucl}
G.~Degrassi, M.~Fedele and P.P.~Giardino, \emph{{Constraints on the trilinear
  Higgs self coupling from precision observables}},
  \href{https://doi.org/10.1007/JHEP04(2017)155}{\emph{JHEP} {\bfseries 04}
  (2017) 155} [\href{https://arxiv.org/abs/1702.01737}{{\ttfamily
  1702.01737}}].

\bibitem{Alasfar:2022zyr}
L.~Alasfar, J.~de~Blas and R.~Gr\"ober, \emph{{Higgs probes of top quark
  contact interactions and their interplay with the Higgs self-coupling}},
  \href{https://doi.org/10.1007/JHEP05(2022)111}{\emph{JHEP} {\bfseries 05}
  (2022) 111} [\href{https://arxiv.org/abs/2202.02333}{{\ttfamily
  2202.02333}}].

\bibitem{ATLAS:2023qzf}
{\scshape ATLAS} collaboration, \emph{{Search for nonresonant pair production
  of Higgs bosons in the bb\textasciimacron{}bb\textasciimacron{} final state
  in pp collisions at s=13\,\,TeV with the ATLAS detector}},
  \href{https://doi.org/10.1103/PhysRevD.108.052003}{\emph{Phys. Rev. D}
  {\bfseries 108} (2023) 052003}
  [\href{https://arxiv.org/abs/2301.03212}{{\ttfamily 2301.03212}}].

\bibitem{ATLAS:2022xzm}
{\scshape ATLAS} collaboration, \emph{{Search for resonant and non-resonant
  Higgs boson pair production in the $ b\overline{b}{\tau}^{+}{\tau}^{-} $
  decay channel using 13 TeV pp collision data from the ATLAS detector}},
  \href{https://doi.org/10.1007/JHEP07(2023)040}{\emph{JHEP} {\bfseries 07}
  (2023) 040} [\href{https://arxiv.org/abs/2209.10910}{{\ttfamily
  2209.10910}}].

\bibitem{ATLAS:2021ifb}
{\scshape ATLAS} collaboration, \emph{{Search for Higgs boson pair production
  in the two bottom quarks plus two photons final state in $pp$ collisions at
  $\sqrt{s}=13$ TeV with the ATLAS detector}},
  \href{https://doi.org/10.1103/PhysRevD.106.052001}{\emph{Phys. Rev. D}
  {\bfseries 106} (2022) 052001}
  [\href{https://arxiv.org/abs/2112.11876}{{\ttfamily 2112.11876}}].

\bibitem{Zabinski:2023jhr}
{\scshape ATLAS} collaboration, \emph{{Probing the nature of electroweak
  symmetry breaking with Higgs boson pairs in ATLAS}},  in \emph{{30th
  International Workshop on Deep-Inelastic Scattering and Related Subjects}},
  7, 2023 [\href{https://arxiv.org/abs/2307.11467}{{\ttfamily 2307.11467}}].

\bibitem{CMS:2022omp}
{\scshape CMS} collaboration, \emph{{Search for nonresonant Higgs boson pair
  production in the four leptons plus twob jets final state in proton-proton
  collisions at $ \sqrt{s} $ = 13 TeV}},
  \href{https://doi.org/10.1007/JHEP06(2023)130}{\emph{JHEP} {\bfseries 06}
  (2023) 130} [\href{https://arxiv.org/abs/2206.10657}{{\ttfamily
  2206.10657}}].

\bibitem{CMS:2020tkr}
{\scshape CMS} collaboration, \emph{{Search for nonresonant Higgs boson pair
  production in final states with two bottom quarks and two photons in
  proton-proton collisions at $ \sqrt{s} $ = 13 TeV}},
  \href{https://doi.org/10.1007/JHEP03(2021)257}{\emph{JHEP} {\bfseries 03}
  (2021) 257} [\href{https://arxiv.org/abs/2011.12373}{{\ttfamily
  2011.12373}}].

\bibitem{CMS:2022hgz}
{\scshape CMS} collaboration, \emph{{Search for nonresonant Higgs boson pair
  production in final state with two bottom quarks and two tau leptons in
  proton-proton collisions at s=13~TeV}},
  \href{https://doi.org/10.1016/j.physletb.2022.137531}{\emph{Phys. Lett. B}
  {\bfseries 842} (2023) 137531}
  [\href{https://arxiv.org/abs/2206.09401}{{\ttfamily 2206.09401}}].

\bibitem{CMS:2022cpr}
{\scshape CMS} collaboration, \emph{{Search for Higgs Boson Pair Production in
  the Four b Quark Final State in Proton-Proton Collisions at s=13\,\,TeV}},
  \href{https://doi.org/10.1103/PhysRevLett.129.081802}{\emph{Phys. Rev. Lett.}
  {\bfseries 129} (2022) 081802}
  [\href{https://arxiv.org/abs/2202.09617}{{\ttfamily 2202.09617}}].

\bibitem{Chang:2018uwu}
J.~Chang, K.~Cheung, J.S.~Lee, C.-T.~Lu and J.~Park, \emph{{Higgs-boson-pair
  production
  H(\textrightarrow{}bb\textasciimacron{})H(\textrightarrow{}\ensuremath{\gamma}\ensuremath{\gamma})
  from gluon fusion at the HL-LHC and HL-100 TeV hadron collider}},
  \href{https://doi.org/10.1103/PhysRevD.100.096001}{\emph{Phys. Rev. D}
  {\bfseries 100} (2019) 096001}
  [\href{https://arxiv.org/abs/1804.07130}{{\ttfamily 1804.07130}}].

\bibitem{ATLAS:2025wdq}
{\scshape ATLAS} collaboration, \emph{{Projected sensitivity of measurements of
  Higgs boson pair production with the ATLAS experiment at the HL-LHC}},
  ATL-PHYS-PUB-2025-006.

\bibitem{He:2015spf}
H.-J.~He, J.~Ren and W.~Yao, \emph{{Probing new physics of cubic Higgs boson
  interaction via Higgs pair production at hadron colliders}},
  \href{https://doi.org/10.1103/PhysRevD.93.015003}{\emph{Phys. Rev. D}
  {\bfseries 93} (2016) 015003}
  [\href{https://arxiv.org/abs/1506.03302}{{\ttfamily 1506.03302}}].

\bibitem{Barr:2013tda}
A.J.~Barr, M.J.~Dolan, C.~Englert and M.~Spannowsky, \emph{{Di-Higgs final
  states augMT2ed -- selecting $hh$ events at the high luminosity LHC}},
  \href{https://doi.org/10.1016/j.physletb.2013.12.011}{\emph{Phys. Lett. B}
  {\bfseries 728} (2014) 308}
  [\href{https://arxiv.org/abs/1309.6318}{{\ttfamily 1309.6318}}].

\bibitem{Goncalves:2018qas}
D.~Gon\c{c}alves, T.~Han, F.~Kling, T.~Plehn and M.~Takeuchi, \emph{{Higgs
  boson pair production at future hadron colliders: From kinematics to
  dynamics}}, \href{https://doi.org/10.1103/PhysRevD.97.113004}{\emph{Phys.
  Rev. D} {\bfseries 97} (2018) 113004}
  [\href{https://arxiv.org/abs/1802.04319}{{\ttfamily 1802.04319}}].

\bibitem{Taliercio:2022maa}
A.~Taliercio, P.~Mastrapasqua, C.~Caputo, P.~Vischia, N.~De~Filippis and
  P.~Bhat, \emph{{Higgs Self Couplings Measurements at Future proton-proton
  Colliders: a Snowmass White Paper}},  in \emph{{Snowmass 2021}}, 3, 2022
  [\href{https://arxiv.org/abs/2203.08042}{{\ttfamily 2203.08042}}].

\bibitem{Braibant:2022ebu}
S.~Braibant, \emph{{Higgs measurements at the Future Circular Colliders}},
  \href{https://doi.org/10.22323/1.398.0616}{\emph{PoS} {\bfseries EPS-HEP2021}
  (2022) 616}.

\bibitem{Mangano:2020sao}
M.L.~Mangano, G.~Ortona and M.~Selvaggi, \emph{{Measuring the Higgs
  self-coupling via Higgs-pair production at a 100 TeV p-p collider}},
  \href{https://doi.org/10.1140/epjc/s10052-020-08595-3}{\emph{Eur. Phys. J. C}
  {\bfseries 80} (2020) 1030}
  [\href{https://arxiv.org/abs/2004.03505}{{\ttfamily 2004.03505}}].

\bibitem{Stapf:2023ndn}
B.~Stapf, A.~Taliercio, E.~Gallo, K.~Tackmann and P.~Mastrapasqua, \emph{{Higgs
  self-coupling measurements at the FCC-hh}},
  \href{https://doi.org/10.22323/1.449.0413}{\emph{PoS} {\bfseries EPS-HEP2023}
  (2024) 413} [\href{https://arxiv.org/abs/2312.03513}{{\ttfamily
  2312.03513}}].

\bibitem{Li:2022zyn}
{\scshape FCC} collaboration, \emph{{Perspectives for Higgs measurements at
  Future Circular Colliders}},
  \href{https://doi.org/10.22323/1.380.0420}{\emph{PoS} {\bfseries PANIC2021}
  (2022) 420}.

\bibitem{Li:2024mrd}
{\scshape FCC} collaboration, \emph{{Higgs physics opportunities at FCC}},
  \href{https://doi.org/10.22323/1.449.0391}{\emph{PoS} {\bfseries EPS-HEP2023}
  (2024) 391}.

\bibitem{Kawada:2022jac}
S.-i.~Kawada, \emph{{ILC Higgs Physics Potential}},
  \href{https://doi.org/10.22323/1.380.0396}{\emph{PoS} {\bfseries PANIC2021}
  (2022) 396}.

\bibitem{List:2024ukv}
J.~List, B.~Bliewert, D.~Ntounis, J.~Tian, C.~Vernieri and J.M.~Torndal,
  \emph{{Higgs Self-coupling Strategy at Linear e+e- Colliders}},
  \href{https://doi.org/10.22323/1.476.0079}{\emph{PoS} {\bfseries ICHEP2024}
  (2025) 079} [\href{https://arxiv.org/abs/2411.01507}{{\ttfamily
  2411.01507}}].

\bibitem{Buonincontri:2022ylv}
{\scshape Muon Collider Physics and Detector Working Group} collaboration,
  \emph{{Higgs boson couplings at muon collider}},
  \href{https://doi.org/10.22323/1.398.0619}{\emph{PoS} {\bfseries EPS-HEP2021}
  (2022) 619}.

\bibitem{Buonincontri:2021okq}
L.~Buonincontri, \emph{{Study of Higgs couplings measurements at muon
  collider}}, \href{https://doi.org/10.1393/ncc/i2021-21031-8}{\emph{Nuovo Cim.
  C} {\bfseries 44} (2021) 31}.

\bibitem{Han:2020pif}
T.~Han, D.~Liu, I.~Low and X.~Wang, \emph{{Electroweak couplings of the Higgs
  boson at a multi-TeV muon collider}},
  \href{https://doi.org/10.1103/PhysRevD.103.013002}{\emph{Phys. Rev. D}
  {\bfseries 103} (2021) 013002}
  [\href{https://arxiv.org/abs/2008.12204}{{\ttfamily 2008.12204}}].

\bibitem{Buonincontri:2024tpa}
L.~Buonincontri, M.~Casarsa, L.~Giambastiani, D.~Lucchesi, L.~Sestini and
  D.~Zuliani, \emph{{Higgs physics at Muon Collider with detailed detector
  simulation}}, \href{https://doi.org/10.1393/ncc/i2024-24288-3}{\emph{Nuovo
  Cim. C} {\bfseries 47} (2024) 288}.

\bibitem{Goertz:2013kp}
F.~Goertz, A.~Papaefstathiou, L.L.~Yang and J.~Zurita, \emph{{Higgs Boson
  self-coupling measurements using ratios of cross sections}},
  \href{https://doi.org/10.1007/JHEP06(2013)016}{\emph{JHEP} {\bfseries 06}
  (2013) 016} [\href{https://arxiv.org/abs/1301.3492}{{\ttfamily 1301.3492}}].

\bibitem{Behr:2015oqq}
J.K.~Behr, D.~Bortoletto, J.A.~Frost, N.P.~Hartland, C.~Issever and J.~Rojo,
  \emph{{Boosting Higgs pair production in the $b\bar{b}b\bar{b}$ final state
  with multivariate techniques}},
  \href{https://doi.org/10.1140/epjc/s10052-016-4215-5}{\emph{Eur. Phys. J. C}
  {\bfseries 76} (2016) 386}
  [\href{https://arxiv.org/abs/1512.08928}{{\ttfamily 1512.08928}}].

\bibitem{Wardrope:2014kya}
D.~Wardrope, E.~Jansen, N.~Konstantinidis, B.~Cooper, R.~Falla and
  N.~Norjoharuddeen, \emph{{Non-resonant Higgs-pair production in the
  $b\overline{b}$ $b\overline{b}$ final state at the LHC}},
  \href{https://doi.org/10.1140/epjc/s10052-015-3439-0}{\emph{Eur. Phys. J. C}
  {\bfseries 75} (2015) 219} [\href{https://arxiv.org/abs/1410.2794}{{\ttfamily
  1410.2794}}].

\bibitem{FerreiradeLima:2014qkf}
D.E.~Ferreira~de Lima, A.~Papaefstathiou and M.~Spannowsky, \emph{{Standard
  model Higgs boson pair production in the $(b\bar{b})(b\bar{b})$ final
  state}}, \href{https://doi.org/10.1007/JHEP08(2014)030}{\emph{JHEP}
  {\bfseries 08} (2014) 030} [\href{https://arxiv.org/abs/1404.7139}{{\ttfamily
  1404.7139}}].

\bibitem{Li:2019tfd}
H.-L.~Li, M.~Ramsey-Musolf and S.~Willocq, \emph{{Probing a scalar
  singlet-catalyzed electroweak phase transition with resonant di-Higgs boson
  production in the $4b$ channel}},
  \href{https://doi.org/10.1103/PhysRevD.100.075035}{\emph{Phys. Rev. D}
  {\bfseries 100} (2019) 075035}
  [\href{https://arxiv.org/abs/1906.05289}{{\ttfamily 1906.05289}}].

\bibitem{Alves:2019igs}
A.~Alves, D.~Gon\c{c}alves, T.~Ghosh, H.-K.~Guo and K.~Sinha, \emph{{Di-Higgs
  Production in the $4b$ Channel and Gravitational Wave Complementarity}},
  \href{https://doi.org/10.1007/JHEP03(2020)053}{\emph{JHEP} {\bfseries 03}
  (2020) 053} [\href{https://arxiv.org/abs/1909.05268}{{\ttfamily
  1909.05268}}].

\bibitem{Amacker:2020bmn}
J.~Amacker et~al., \emph{{Higgs self-coupling measurements using deep learning
  in the $ b\overline{b}b\overline{b} $ final state}},
  \href{https://doi.org/10.1007/JHEP12(2020)115}{\emph{JHEP} {\bfseries 12}
  (2020) 115} [\href{https://arxiv.org/abs/2004.04240}{{\ttfamily
  2004.04240}}].

\bibitem{Chiang:2024pho}
C.-W.~Chiang, F.-Y.~Hsieh, S.-C.~Hsu and I.~Low, \emph{{Deep learning to
  improve the sensitivity of Di-Higgs searches in the 4b channel}},
  \href{https://doi.org/10.1007/JHEP09(2024)139}{\emph{JHEP} {\bfseries 09}
  (2024) 139} [\href{https://arxiv.org/abs/2401.14198}{{\ttfamily
  2401.14198}}].

\bibitem{Alison:2019kud}
J.~Alison, S.~An, P.~Bryant, B.~Burkle, S.~Gleyzer, M.~Narain et~al.,
  \emph{{End-to-end particle and event identification at the Large Hadron
  Collider with CMS Open Data}},  in \emph{{Meeting of the Division of
  Particles and Fields of the American Physical Society}}, 10, 2019
  [\href{https://arxiv.org/abs/1910.07029}{{\ttfamily 1910.07029}}].

\bibitem{Andrews:2019faz}
M.~Andrews, J.~Alison, S.~An, P.~Bryant, B.~Burkle, S.~Gleyzer et~al.,
  \emph{{End-to-end jet classification of quarks and gluons with the CMS Open
  Data}}, \href{https://doi.org/10.1016/j.nima.2020.164304}{\emph{Nucl.
  Instrum. Meth. A} {\bfseries 977} (2020) 164304}
  [\href{https://arxiv.org/abs/1902.08276}{{\ttfamily 1902.08276}}].

\bibitem{Andrews:2018nwy}
M.~Andrews, M.~Paulini, S.~Gleyzer and B.~Poczos, \emph{{End-to-End Physics
  Event Classification with CMS Open Data: Applying Image-Based Deep Learning
  to Detector Data for the Direct Classification of Collision Events at the
  LHC}}, \href{https://doi.org/10.1007/s41781-020-00038-8}{\emph{Comput. Softw.
  Big Sci.} {\bfseries 4} (2020) 6}
  [\href{https://arxiv.org/abs/1807.11916}{{\ttfamily 1807.11916}}].

\bibitem{Baldi:2014kfa}
P.~Baldi, P.~Sadowski and D.~Whiteson, \emph{{Searching for Exotic Particles in
  High-Energy Physics with Deep Learning}},
  \href{https://doi.org/10.1038/ncomms5308}{\emph{Nature Commun.} {\bfseries 5}
  (2014) 4308} [\href{https://arxiv.org/abs/1402.4735}{{\ttfamily 1402.4735}}].

\bibitem{Abdughani:2019wuv}
M.~Abdughani, J.~Ren, L.~Wu, J.M.~Yang and J.~Zhao, \emph{{Supervised deep
  learning in high energy phenomenology: a mini review}},
  \href{https://doi.org/10.1088/0253-6102/71/8/955}{\emph{Commun. Theor. Phys.}
  {\bfseries 71} (2019) 955}
  [\href{https://arxiv.org/abs/1905.06047}{{\ttfamily 1905.06047}}].

\bibitem{Guest:2016iqz}
D.~Guest, J.~Collado, P.~Baldi, S.-C.~Hsu, G.~Urban and D.~Whiteson, \emph{{Jet
  Flavor Classification in High-Energy Physics with Deep Neural Networks}},
  \href{https://doi.org/10.1103/PhysRevD.94.112002}{\emph{Phys. Rev. D}
  {\bfseries 94} (2016) 112002}
  [\href{https://arxiv.org/abs/1607.08633}{{\ttfamily 1607.08633}}].

\bibitem{Kasieczka:2019dbj}
A.~Butter et~al., \emph{{The Machine Learning landscape of top taggers}},
  \href{https://doi.org/10.21468/SciPostPhys.7.1.014}{\emph{SciPost Phys.}
  {\bfseries 7} (2019) 014} [\href{https://arxiv.org/abs/1902.09914}{{\ttfamily
  1902.09914}}].

\bibitem{Woodward:2024dxb}
N.S.~Woodward, S.E.~Park, G.~Grosso, J.~Krupa and P.~Harris, \emph{{Product
  Manifold Machine Learning for Physics}},
  \href{https://arxiv.org/abs/2412.07033}{{\ttfamily 2412.07033}}.

\bibitem{Lin:2018cin}
J.~Lin, M.~Freytsis, I.~Moult and B.~Nachman, \emph{{Boosting $H\to b\bar b$
  with Machine Learning}},
  \href{https://doi.org/10.1007/JHEP10(2018)101}{\emph{JHEP} {\bfseries 10}
  (2018) 101} [\href{https://arxiv.org/abs/1807.10768}{{\ttfamily
  1807.10768}}].

\bibitem{Cogan:2014oua}
J.~Cogan, M.~Kagan, E.~Strauss and A.~Schwarztman, \emph{{Jet-Images: Computer
  Vision Inspired Techniques for Jet Tagging}},
  \href{https://doi.org/10.1007/JHEP02(2015)118}{\emph{JHEP} {\bfseries 02}
  (2015) 118} [\href{https://arxiv.org/abs/1407.5675}{{\ttfamily 1407.5675}}].

\bibitem{deOliveira:2015xxd}
L.~de~Oliveira, M.~Kagan, L.~Mackey, B.~Nachman and A.~Schwartzman,
  \emph{{Jet-images \textemdash{} deep learning edition}},
  \href{https://doi.org/10.1007/JHEP07(2016)069}{\emph{JHEP} {\bfseries 07}
  (2016) 069} [\href{https://arxiv.org/abs/1511.05190}{{\ttfamily
  1511.05190}}].

\bibitem{Lee:2019cad}
J.S.H.~Lee, I.~Park, I.J.~Watson and S.~Yang, \emph{{Quark-Gluon Jet
  Discrimination Using Convolutional Neural Networks}},
  \href{https://doi.org/10.3938/jkps.74.219}{\emph{J. Korean Phys. Soc.}
  {\bfseries 74} (2019) 219}
  [\href{https://arxiv.org/abs/2012.02531}{{\ttfamily 2012.02531}}].

\bibitem{Madrazo:2017qgh}
C.F.~Madrazo, I.H.~Cacha, L.L.~Iglesias and J.M.~de~Lucas, \emph{{Application
  of a Convolutional Neural Network for image classification for the analysis
  of collisions in High Energy Physics}},
  \href{https://doi.org/10.1051/epjconf/201921406017}{\emph{EPJ Web Conf.}
  {\bfseries 214} (2019) 06017}
  [\href{https://arxiv.org/abs/1708.07034}{{\ttfamily 1708.07034}}].

\bibitem{Dreyer:2020brq}
F.A.~Dreyer and H.~Qu, \emph{{Jet tagging in the Lund plane with graph
  networks}}, \href{https://doi.org/10.1007/JHEP03(2021)052}{\emph{JHEP}
  {\bfseries 03} (2021) 052}
  [\href{https://arxiv.org/abs/2012.08526}{{\ttfamily 2012.08526}}].

\bibitem{Abdughani:2018wrw}
M.~Abdughani, J.~Ren, L.~Wu and J.M.~Yang, \emph{{Probing stop pair production
  at the LHC with graph neural networks}},
  \href{https://doi.org/10.1007/JHEP08(2019)055}{\emph{JHEP} {\bfseries 08}
  (2019) 055} [\href{https://arxiv.org/abs/1807.09088}{{\ttfamily
  1807.09088}}].

\bibitem{Gong:2022lye}
S.~Gong, Q.~Meng, J.~Zhang, H.~Qu, C.~Li, S.~Qian et~al., \emph{{An efficient
  Lorentz equivariant graph neural network for jet tagging}},
  \href{https://doi.org/10.1007/JHEP07(2022)030}{\emph{JHEP} {\bfseries 07}
  (2022) 030} [\href{https://arxiv.org/abs/2201.08187}{{\ttfamily
  2201.08187}}].

\bibitem{Qu:2019gqs}
H.~Qu and L.~Gouskos, \emph{{ParticleNet: Jet Tagging via Particle Clouds}},
  \href{https://doi.org/10.1103/PhysRevD.101.056019}{\emph{Phys. Rev. D}
  {\bfseries 101} (2020) 056019}
  [\href{https://arxiv.org/abs/1902.08570}{{\ttfamily 1902.08570}}].

\bibitem{Mikuni:2021pou}
V.~Mikuni and F.~Canelli, \emph{{Point cloud transformers applied to collider
  physics}}, \href{https://doi.org/10.1088/2632-2153/ac07f6}{\emph{Mach. Learn.
  Sci. Tech.} {\bfseries 2} (2021) 035027}
  [\href{https://arxiv.org/abs/2102.05073}{{\ttfamily 2102.05073}}].

\bibitem{Komiske:2018cqr}
P.T.~Komiske, E.M.~Metodiev and J.~Thaler, \emph{{Energy Flow Networks: Deep
  Sets for Particle Jets}},
  \href{https://doi.org/10.1007/JHEP01(2019)121}{\emph{JHEP} {\bfseries 01}
  (2019) 121} [\href{https://arxiv.org/abs/1810.05165}{{\ttfamily
  1810.05165}}].

\bibitem{Usman:2024hxz}
M.~Usman, M.H.~Shahid, M.~Ejaz, U.~Hani, N.~Fatima, A.R.~Khan et~al.,
  \emph{{Particle Multi-Axis Transformer for Jet Tagging}},
  \href{https://arxiv.org/abs/2406.06638}{{\ttfamily 2406.06638}}.

\bibitem{Qu:2022mxj}
H.~Qu, C.~Li and S.~Qian, \emph{{Particle Transformer for Jet Tagging}},
  \href{https://arxiv.org/abs/2202.03772}{{\ttfamily 2202.03772}}.

\bibitem{Tagami:2024gtc}
R.~Tagami, T.~Suehara and M.~Ishino, \emph{{Application of Particle Transformer
  to quark flavor tagging in the ILC project}},
  \href{https://doi.org/10.1051/epjconf/202431503011}{\emph{EPJ Web Conf.}
  {\bfseries 315} (2024) 03011}
  [\href{https://arxiv.org/abs/2410.11322}{{\ttfamily 2410.11322}}].

\bibitem{Camagni:2024zzi}
{\scshape CMS} collaboration, \emph{{Particle Transformer for $\tau$ lepton
  pair invariant mass reconstruction for the $HH\to b \bar{b} \tau^+\tau^-$ CMS
  analysis}}, \href{https://doi.org/10.1393/ncc/i2024-24250-5}{\emph{Nuovo Cim.
  C} {\bfseries 47} (2024) 250}.

\bibitem{Wu:2024thh}
Y.~Wu, K.~Wang, C.~Li, H.~Qu and J.~Zhu, \emph{{Jet tagging with
  more-interaction particle transformer*}},
  \href{https://doi.org/10.1088/1674-1137/ad7f3d}{\emph{Chin. Phys. C}
  {\bfseries 49} (2025) 013110}
  [\href{https://arxiv.org/abs/2407.08682}{{\ttfamily 2407.08682}}].

\bibitem{He:2023cfc}
M.~He and D.~Wang, \emph{{Quark/gluon discrimination and top tagging with dual
  attention transformer}},
  \href{https://doi.org/10.1140/epjc/s10052-023-12293-1}{\emph{Eur. Phys. J. C}
  {\bfseries 83} (2023) 1116}
  [\href{https://arxiv.org/abs/2307.04723}{{\ttfamily 2307.04723}}].

\bibitem{Apresyan:2022tqw}
A.~Apresyan et~al., \emph{{Improving Di-Higgs Sensitivity at Future Colliders
  in Hadronic Final States with Machine Learning}},  in \emph{{Snowmass 2021}},
  3, 2022 [\href{https://arxiv.org/abs/2203.07353}{{\ttfamily 2203.07353}}].

\bibitem{Stylianou:2023xit}
P.~Stylianou and G.~Weiglein, \emph{{Constraints on the trilinear and quartic
  Higgs couplings from triple Higgs production at the LHC and beyond}},
  \href{https://doi.org/10.1140/epjc/s10052-024-12722-9}{\emph{Eur. Phys. J. C}
  {\bfseries 84} (2024) 366}
  [\href{https://arxiv.org/abs/2312.04646}{{\ttfamily 2312.04646}}].

\bibitem{Alasfar:2022vqw}
L.~Alasfar, R.~Gr\"ober, C.~Grojean, A.~Paul and Z.~Qian, \emph{{Machine
  learning the trilinear and light-quark Yukawa couplings from Higgs pair
  kinematic shapes}},
  \href{https://doi.org/10.1007/JHEP11(2022)045}{\emph{JHEP} {\bfseries 11}
  (2022) 045} [\href{https://arxiv.org/abs/2207.04157}{{\ttfamily
  2207.04157}}].

\bibitem{Huang:2022rne}
L.~Huang, S.-b.~Kang, J.H.~Kim, K.~Kong and J.S.~Pi, \emph{{Portraying double
  Higgs at the Large Hadron Collider II}},
  \href{https://doi.org/10.1007/JHEP08(2022)114}{\emph{JHEP} {\bfseries 08}
  (2022) 114} [\href{https://arxiv.org/abs/2203.11951}{{\ttfamily
  2203.11951}}].

\bibitem{Tannenwald:2020mhq}
B.~Tannenwald, C.~Neu, A.~Li, G.~Buehlmann, A.~Cuddeback, L.~Hatfield et~al.,
  \emph{{Benchmarking Machine Learning Techniques with Di-Higgs Production at
  the LHC}},  \href{https://arxiv.org/abs/2009.06754}{{\ttfamily 2009.06754}}.

\bibitem{Abdughani:2020xfo}
M.~Abdughani, D.~Wang, L.~Wu, J.M.~Yang and J.~Zhao, \emph{{Probing the triple
  Higgs boson coupling with machine learning at the LHC}},
  \href{https://doi.org/10.1103/PhysRevD.104.056003}{\emph{Phys. Rev. D}
  {\bfseries 104} (2021) 056003}
  [\href{https://arxiv.org/abs/2005.11086}{{\ttfamily 2005.11086}}].

\bibitem{Kerner:2024dgm}
M.~Kerner, \emph{{Higgs Self-Coupling and Yukawa Corrections to Higgs Boson
  Pair Production}}, \href{https://doi.org/10.22323/1.467.0016}{\emph{PoS}
  {\bfseries LL2024} (2024) 016}.

\bibitem{Li:2024iio}
H.T.~Li, Z.-G.~Si, J.~Wang, X.~Zhang and D.~Zhao, \emph{{Improved constraints
  on Higgs boson self-couplings with quartic and cubic power dependencies of
  the cross section*}},
  \href{https://doi.org/10.1088/1674-1137/ad9d1d}{\emph{Chin. Phys. C}
  {\bfseries 49} (2025) 023107}
  [\href{https://arxiv.org/abs/2407.14716}{{\ttfamily 2407.14716}}].

\bibitem{Frederix:2014hta}
R.~Frederix, S.~Frixione, V.~Hirschi, F.~Maltoni, O.~Mattelaer, P.~Torrielli
  et~al., \emph{{Higgs pair production at the LHC with NLO and parton-shower
  effects}}, \href{https://doi.org/10.1016/j.physletb.2014.03.026}{\emph{Phys.
  Lett. B} {\bfseries 732} (2014) 142}
  [\href{https://arxiv.org/abs/1401.7340}{{\ttfamily 1401.7340}}].

\bibitem{Heinrich:2019bkc}
G.~Heinrich, S.P.~Jones, M.~Kerner, G.~Luisoni and L.~Scyboz, \emph{{Probing
  the trilinear Higgs boson coupling in di-Higgs production at NLO QCD
  including parton shower effects}},
  \href{https://doi.org/10.1007/JHEP06(2019)066}{\emph{JHEP} {\bfseries 06}
  (2019) 066} [\href{https://arxiv.org/abs/1903.08137}{{\ttfamily
  1903.08137}}].

\bibitem{deFlorian:2013jea}
D.~de~Florian and J.~Mazzitelli, \emph{{Higgs Boson Pair Production at
  Next-to-Next-to-Leading Order in QCD}},
  \href{https://doi.org/10.1103/PhysRevLett.111.201801}{\emph{Phys. Rev. Lett.}
  {\bfseries 111} (2013) 201801}
  [\href{https://arxiv.org/abs/1309.6594}{{\ttfamily 1309.6594}}].

\bibitem{deFlorian:2015moa}
D.~de~Florian and J.~Mazzitelli, \emph{{Higgs pair production at
  next-to-next-to-leading logarithmic accuracy at the LHC}},
  \href{https://doi.org/10.1007/JHEP09(2015)053}{\emph{JHEP} {\bfseries 09}
  (2015) 053} [\href{https://arxiv.org/abs/1505.07122}{{\ttfamily
  1505.07122}}].

\bibitem{deFlorian:2013uza}
D.~de~Florian and J.~Mazzitelli, \emph{{Two-loop virtual corrections to Higgs
  pair production}},
  \href{https://doi.org/10.1016/j.physletb.2013.06.046}{\emph{Phys. Lett. B}
  {\bfseries 724} (2013) 306}
  [\href{https://arxiv.org/abs/1305.5206}{{\ttfamily 1305.5206}}].

\bibitem{Alwall:2014hca}
J.~Alwall, R.~Frederix, S.~Frixione, V.~Hirschi, F.~Maltoni, O.~Mattelaer
  et~al., \emph{{The automated computation of tree-level and next-to-leading
  order differential cross sections, and their matching to parton shower
  simulations}}, \href{https://doi.org/10.1007/JHEP07(2014)079}{\emph{JHEP}
  {\bfseries 07} (2014) 079} [\href{https://arxiv.org/abs/1405.0301}{{\ttfamily
  1405.0301}}].

\bibitem{Sjostrand:2014zea}
T.~Sj\"ostrand, S.~Ask, J.R.~Christiansen, R.~Corke, N.~Desai, P.~Ilten et~al.,
  \emph{{An introduction to PYTHIA 8.2}},
  \href{https://doi.org/10.1016/j.cpc.2015.01.024}{\emph{Comput. Phys. Commun.}
  {\bfseries 191} (2015) 159}
  [\href{https://arxiv.org/abs/1410.3012}{{\ttfamily 1410.3012}}].

\bibitem{deFavereau:2013fsa}
{\scshape DELPHES 3} collaboration, \emph{{DELPHES 3, A modular framework for
  fast simulation of a generic collider experiment}},
  \href{https://doi.org/10.1007/JHEP02(2014)057}{\emph{JHEP} {\bfseries 02}
  (2014) 057} [\href{https://arxiv.org/abs/1307.6346}{{\ttfamily 1307.6346}}].

\bibitem{Cacciari:2011ma}
M.~Cacciari, G.P.~Salam and G.~Soyez, \emph{{FastJet User Manual}},
  \href{https://doi.org/10.1140/epjc/s10052-012-1896-2}{\emph{Eur. Phys. J. C}
  {\bfseries 72} (2012) 1896}
  [\href{https://arxiv.org/abs/1111.6097}{{\ttfamily 1111.6097}}].

\bibitem{Cacciari:2008gp}
M.~Cacciari, G.P.~Salam and G.~Soyez, \emph{{The anti-$k_t$ jet clustering
  algorithm}}, \href{https://doi.org/10.1088/1126-6708/2008/04/063}{\emph{JHEP}
  {\bfseries 04} (2008) 063} [\href{https://arxiv.org/abs/0802.1189}{{\ttfamily
  0802.1189}}].

\bibitem{vaswani2023attentionneed}
A.~Vaswani, N.~Shazeer, N.~Parmar, J.~Uszkoreit, L.~Jones, A.N.~Gomez et~al.,
  \emph{Attention is all you need},
  \href{https://arxiv.org/abs/1706.03762}{{\ttfamily 1706.03762}}.

\bibitem{Fenton:2020woz}
M.J.~Fenton, A.~Shmakov, T.-W.~Ho, S.-C.~Hsu, D.~Whiteson and P.~Baldi,
  \emph{{Permutationless many-jet event reconstruction with symmetry preserving
  attention networks}},
  \href{https://doi.org/10.1103/PhysRevD.105.112008}{\emph{Phys. Rev. D}
  {\bfseries 105} (2022) 112008}
  [\href{https://arxiv.org/abs/2010.09206}{{\ttfamily 2010.09206}}].

\bibitem{Shmakov:2021qdz}
A.~Shmakov, M.J.~Fenton, T.-W.~Ho, S.-C.~Hsu, D.~Whiteson and P.~Baldi,
  \emph{{SPANet: Generalized permutationless set assignment for particle
  physics using symmetry preserving attention}},
  \href{https://doi.org/10.21468/SciPostPhys.12.5.178}{\emph{SciPost Phys.}
  {\bfseries 12} (2022) 178}
  [\href{https://arxiv.org/abs/2106.03898}{{\ttfamily 2106.03898}}].

\bibitem{Fenton:2023ikr}
M.J.~Fenton, A.~Shmakov, H.~Okawa, Y.~Li, K.-Y.~Hsiao, S.-C.~Hsu et~al.,
  \emph{{Reconstruction of unstable heavy particles using deep
  symmetry-preserving attention networks}},
  \href{https://doi.org/10.1038/s42005-024-01627-4}{\emph{Commun. Phys.}
  {\bfseries 7} (2024) 139} [\href{https://arxiv.org/abs/2309.01886}{{\ttfamily
  2309.01886}}].

\bibitem{Wang:2024rup}
A.~Wang, A.~Gandrakota, J.~Ngadiuba, V.~Sahu, P.~Bhatnagar, E.E.~Khoda et~al.,
  \emph{{Interpreting Transformers for Jet Tagging}},  12, 2024
  [\href{https://arxiv.org/abs/2412.03673}{{\ttfamily 2412.03673}}].

\end{thebibliography}\endgroup

\end{document}